\documentclass[fleqn,usenatbib]{mnras}

\usepackage[T1]{fontenc}
\usepackage{ae,aecompl}

\usepackage{graphicx}   
\usepackage{amsmath}    
\usepackage{amssymb}    

\usepackage[pdf]{graphviz}
\usepackage[english]{babel}
\usepackage{xspace}
\usepackage[multi-part-units=single]{siunitx}

\usepackage{ifthen}
\newboolean{isdraft}
\setboolean{isdraft}{false}

\ifthenelse{\boolean{isdraft}}{\usepackage{todonotes}}{}

\DeclareSIUnit[number-unit-product = {}]\yr{yr}
\DeclareSIUnit[number-unit-product = {}]\Mpc{Mpc}
\DeclareSIUnit[number-unit-product = {}]\cMpc{cMpc}
\DeclareSIUnit[number-unit-product = {}]\ckpc{ckpc}
\DeclareSIUnit[number-unit-product = {}]\Mpch{\si{\per\h\mega\pc}}
\DeclareSIUnit[number-unit-product = {}]\kpch{\si{\per\h\kilo\pc}}
\DeclareSIUnit[number-unit-product = {}]\cMpch{\si{\per\h\cMpc}}
\DeclareSIUnit[number-unit-product = {}]\ckpch{\si{\per\h\ckpc}}
\DeclareSIUnit[number-unit-product = {}]\pc{pc}
\DeclareSIUnit[number-unit-product = {}]\h{\textit{h}}
\DeclareSIUnit[number-unit-product = {}]\c{c}
\DeclareSIUnit[number-unit-product = {}]\dex{dex}
\DeclareSIUnit[number-unit-product = {}]\Msol{{M_\odot}}
\DeclareSIUnit[number-unit-product = {}]\Msolh{\si{\per\h\Msol}}

\renewcommand{\v}[1]{\boldsymbol #1}
\newcommand{\rt}{\ensuremath{\rho_{\mathrm thresh}}\xspace}

\newcommand{\rl}{\ensuremath{\rho_{\mathrm lim}}\xspace}
\newcommand{\rper}{\ensuremath{\rho_{\mathrm perc}}\xspace}
\newcommand{\rmean}{\ensuremath{\left<\rho\right>}\xspace}

\newcommand{\ttt}[1]{\ensuremath{10^{#1}}\xspace}
\newcommand{\nttt}[2]{\ensuremath{#1 \cdot 10^{#2}}\xspace}

\newcommand{\method}{Tessellation-Level-Tree\xspace}
\newcommand{\methabb}{\textsc{TLT}\xspace}

\newcommand{\Mpch}{\si{\Mpch}\xspace}

\newcommand{\scp}{\textsuperscript{$+$}}
\newcommand{\scm}{\textsuperscript{$-$}}

\title[Density Peak Hierarchy and Clustering]{The \method: characterising the nested hierarchy of density peaks and their spatial distribution in cosmological N-body simulations}

\author[P. Busch and S. D. M. White]{
Philipp Busch$^{1}$\thanks{E-mail: pbusch@mpa-garching.mpg.de}
and Simon D. M. White$^{1}$
\\
$^{1}$Max-Planck-Institut f\"ur Astrophysik, Postfach 1317, D-85741 Garching, Germany
}

\date{Accepted XXX. Received YYY; in original form ZZZ}

\pubyear{2019}

\begin{document}
\label{firstpage}
\pagerange{\pageref{firstpage}--\pageref{lastpage}}
\maketitle

\begin{abstract}
 We use the Millennium and Millennium-II simulations to illustrate the \method (\methabb), a hierarchical tree structure linking density peaks in a field constructed by voronoi tessellation of the particles in a cosmological N-body simulation.  The \methabb uniquely partitions the simulation particles into disjoint subsets, each associated with a local density peak. Each peak is a subpeak of a unique higher peak. The \methabb can be persistence filtered to suppress peaks produced by discreteness noise. Thresholding a peak's particle list at $\sim 80\rmean$ results in a structure similar to a standard friend-of-friends halo and its subhaloes. For thresholds below $\sim 7\rmean$, the largest structure percolates and is much more massive than other objects. It may be considered as defining the cosmic web. For a threshold of $5\rmean$, it contains about half of all cosmic mass and occupies $\sim 1\%$ of all cosmic volume; a typical external point is then $\sim 7h^{-1}$~Mpc from the web. We investigate the internal structure and clustering of \methabb peaks. Defining the saddle point density \rl as the density at which a peak joins its parent peak, we show the median value of \rl for FoF-like peaks to be similar to the density threshold at percolation. Assembly bias as a function of \rl is stronger than for any known internal halo property. For peaks of group mass and below, the lowest quintile in \rl has $b\approx 0$, and is thus uncorrelated with the mass distribution.

\end{abstract}

\begin{keywords}
large-scale structure of Universe -- methods: data analysis
\end{keywords}



\section{Introduction}


In the current standard paradigm for cosmological structure formation, the concordance $\mathrm{\Lambda CDM}$ model, cold dark matter (CDM) dominates the cosmic mass budget and gravity drives structural evolution from the low-amplitude, gaussian fluctuation field visible in the cosmic microwave background radiation to today's highly structured, nonlinear network, the cosmic web \citep{shandarin_large-scale_1989,bond_how_1996}. At late times this evolution occurs within a universe where the expansion is being accelerated by dark energy in the form of an effective cosmological constant, hence the $\mathrm{\Lambda}$ in $\mathrm{\Lambda CDM}$. The cosmic web is built of overdense filaments and sheets which link dense, centrally concentrated structures called haloes. These form through anisotropic gravitational collapse and are the birth-places and current hosts of galaxies \citep{white_core_1978}. In the inner regions of haloes, dark matter densities reach values exceeding the mean by many orders of magnitude \citep[e.g.][]{pandey_exploring_2013}. 

This hierarchy of structures, subhaloes embedded in larger haloes which are in turn embedded in the cosmic web, is usually investigated with the help of cosmological simulations \citep[see][for reviews]{bagla_cosmological_2005,trenti_gravitational_2008,frenk_dark_2012}. In recent years such simulations have increasingly included  hydrodynamical modelling in order to treat the evolution of the baryonic components in addition to that of the dark matter \citep{schaye_eagle_2015, vogelsberger_introducing_2014, khandai_massiveblack-ii_2015, dubois_horizon-agn_2016,pillepich_simulating_2018}. A wide variety of algorithms have been used to identify galaxies, galaxy clusters and the cosmic web within such simulations. In particular,
since dark matter haloes play such a central role, a large number of halo-finders have been developed. While all have the same goal, they differ significantly in approach; the intrinsic complexity of cosmic structure results in each identifying a halo population with somewhat different characteristics. For example, two of the oldest and most basic halo-finders are the friends-of-friends (FoF) \citep{davis_evolution_1985} and spherical overdensity (SO) \citep{lacey_merger_1994} algorithms. The former often links almost disjoint haloes with low-density bridges which may sometimes reflect discreteness noise rather the true cosmic web. Such composite ``haloes" are much less prominent in catalogues constructed with the SO algorithm, but these are geometrically biased by the spherical boundary which it imposes.

Most more modern halo finders explicitly address halo complexity by attempting to identify all subhaloes within each halo, where a subhalo is defined to contain a single significant local density peak. Subhaloes may defined in 3D configuration space, as in algorithms such as \textsc{subfind} \citep{springel_populating_2001} and \textsc{adaptahop} \citep{aubert_origin_2004} and in the \method  (\methabb) studied here, or in 6D phase space as in \textsc{rockstar} \citep{behroozi_rockstar_2013}. These algorithms are often supplemented by additional criteria such as requiring subhaloes to be gravitationally self-bound \citep[e.g.][]{springel_populating_2001, behroozi_rockstar_2013} or temporally persistent \citep[e.g.][]{han_resolving_2012,han_hbt+:_2018}. A more complete discussion of these issues and others 
can be found in \cite{knebe_structure_2013}.

Tessellation methods have long been used to estimate cosmological density fields, dating back at least to \cite{van_de_weygaert_fragmenting_1994}. Their advantages are that they are uniquely defined by the positions of the given objects (e.g. galaxies, or simulation particles), that they are fully spatially adaptive and free from geometric assumptions, thus can deal with large density variations and complex or irregular structures, and that they require no smoothing beyond that implied by the discreteness of the point set. An excellent overview of the properties of Voronoi and Delaunay tessellations as applied to cosmological large-scale structure can be found in the lecture notes of \cite{van_de_weygaert_cosmic_2009}, while a more mathematical discussion in a more general context can be found in the textbook of \cite{okabe_spatial_2000}.


A halo finder which has many commonalities with our \methabb is \textsc{voboz} \citep{neyrinck_voboz_2005} which uses a Voronoi tessellation of the simulated particle
distribution to estimate a density field, which is taken to be uniform within each cell and inversely proportional to its volume. \textsc{voboz} then identifies local density peaks and binds to them all mutually neighboring cells that lie above a
user-selected density threshold.\footnote{An equivalent void identification algorithm called \textsc{zobov} \citep{neyrinck_zobov_2008} has been very successful and is part of the \textsc{vide} \citep{sutter_vide_2015} void identification toolkit. This technique is closely related to the watershed void finder of \cite{platen_cosmic_2007} which is based on the DTFE density field.}
Most applications of halo-finders only consider objects defined above a density threshold which is far above the typical density of the cosmic web, thereby clearly distinguishing haloes from their environment. While this is sufficient to characterise the haloes themselves, and allows quantification of certain
aspects of their large-scale spatial distribution \citep[e.g.][]{sheth_large-scale_1999}, it is neither able, nor tries, to capture the morphology of the transition to the cosmic web or of the web itself.


Just as for halo-finders, there is a plethora of algorithms which identify variously defined versions of the cosmic web (see \cite{libeskind_tracing_2018} for a discussion). While the detailed definitions differ, the primary aim of all of these algorithms is to define a space-filling filamentary network from the matter density field. In many cases, the classification also extends to find nodes of the network, as well as walls spanning between filaments, and voids surrounded by the network. Unlike the \methabb we present below, these algorithms typically use a subsample of the simulation particles and/or a gridded or otherwise smoothed density field.

A particularly elegant example of such methods is the  \mbox{DisPerSE} formalism presented in \cite{sousbie_persistentI_2011} and \cite{sousbie_persistentII_2011}, which uses discrete Morse theory to identify a network of density ridges which traces the cosmic web. Further related material can be found in the comprehensive studies of van de Weygaert and collaborators who, in recent years, have applied (computational) topology to analyse a variety of aspects of structure in cosmological simulations \citep{van_de_weygaert_alpha_2011,pranav_topology_2017,feldbrugge_stochastic_2019}. The textbook by \cite{edelsbrunner_computational_2010} provides an excellent overview of the general topic of computational topology.

Knowledge of the cosmic web and its morphology has two main uses. On small scales it allows the environment of galaxies to be characterised, and thus furthers our understanding of the interplay between environment and galaxy formation from both observational \citep[eg.][]{kraljic_galaxy_2018} and theoretical \citep[eg.][]{borzyszkowski_zomg_2017} points of view. On larger scales quantifying the morphology of the web may give information about cosmological parameters and the initial conditions for structure formation \citep[see, e.g.][]{shim_massive_2014,lee_observational_2015,massara_voids_2015,kreisch_massive_2019}. 

Perhaps the simplest way to characterise the connection between halo properties and the larger scale environment is through the dependence they induce in the two-point statistics of halo clustering. It has long been known that the strength of halo
clustering increases with halo mass \citep[see e.g.][]{kaiser_spatial_1984, mo_analytic_1996} so that haloes are biased tracers of the underlying matter distribution. More recent work has shown, however, that halo clustering depends not only on mass, but also on additional structural properties such as concentration, shape, spin velocity anisotropy and substructure content \citep{gao_age_2005,wechsler_dependence_2006,gao_assembly_2007,dalal_halo_2008,faltenbacher_assembly_2010,lazeyras_large-scale_2017}. Because such properties depend on how haloes are assembled, this effect has become known as \emph{halo assembly bias}.

This paper aims to link haloes to the cosmic web and to larger scale structure by tracing the connectivity of the density field from the highest to the lowest densities in N-body simulations. For this we estimate each particle's density using the volume of its Voronoi cell, and we group particles into objects by connecting them over shared Voronoi cell faces. We begin by describing 
how this enables us to put all particles into a hierarchical tree structure, the \method (\methabb), which uniquely connects each particle to a local peak and each peak to a higher peak of which it can be considered a subpeak. The method itself is presented in \autoref{sec:method} where we also explain how the peak list can be persistence filtered to exclude noise peaks and thresholded to produce halo/subhalo catalogues similar to those of standard group-finders. We then use the \methabb to investigate the percolation of density level surfaces and the properties of the cosmic web, which we define as the largest connected object above a percolating level surface \autoref{sec:above}. In the following section (\autoref{sec:abundance}) we study halo abundance as a function of limiting density, and compare with the result found using a standard FoF algorithm. Next we briefly showcase two applications of the \methabb that will be discussed in follow-up papers: the study of the mass-density distribution within haloes as a function of their mass (\autoref{sec:mass_dens_profs}) and the remarkably strong assembly bias signal that we find as a function of the saddle-point density at which a halo is connected to a larger object (\autoref{sec:ab_res}). Finally, we summarise our conclusions.

\section{Methodology}\label{sec:method}
  
  The first part of this section (\autoref{sec:the_method}) describes the construction of the \method and defines a number of peak properties derived from it. It ends with a quick review of the terminology used for objects in the remainder of our paper. Some of the peaks in the raw \methabb
  can be considered as reflecting discreteness noise in the way the simulation particles sample an (assumed smooth) underlying density field. In \autoref{sec:pers} we consider both Voronoi and Delaunay tesselations as possible bases for estimating the density local to each particle, finding the former to be significantly lower noise for our purposes, as judged by tests on Poisson-distributed point distributions. We then define a persistence criterion to filter out the peaks most likely to be ``spurious".  In \autoref{sec:halo_def} we describe how the \methabb can be used to catalogue a variety of objects, some more, some less similar to the haloes and subhaloes defined by standard group-finders. In \autoref{sec:deriv_quant} we then define various properties of these objects that can easily be computed using the \methabb. 
  
  \subsection{The \method}\label{sec:the_method}

      The \method (\methabb) is a structure defined on a spatial distribution of points, for each of which a density value and a neighbour list is given. In this paper the points will always be the 3D positions of the full particle set of an N-body simulation. The algorithm finds all density peaks, defined as particles which are denser than all their neighbours. It then partitions the full particle list into disjoint subsets, one associated with each peak, and orders the points by density within each subset. Finally, each peak (except the highest) is linked to a higher peak of which it can be considered a subpeak. Particles assigned to a particular peak all have density higher than that at the highest saddle point linking that peak to a higher peak and so can be considered as the particle set surrounding the peak and enclosed by a level surface at this saddle-point density. The resulting tree structure links all particles in the simulation and is unique as long as there are no duplicates in the set of density values. 
      
      Throughout this paper, except briefly in  \autoref{sec:pers}, we estimate densities for each particle using a Voronoi tessellation generated from the particle positions. The inverse of the volume of the Voronoi cell of each particle  serves as our density estimate. The faces of the cells give us the connectivity on this unstructured grid, in the form of a neighbour list for each particle. The tessellation is performed by the routines of the \verb+AREPO+ code \citep{springel_e_2010} which are efficient enough to tessellate all $10^{10}$ particles of the Millennium Simulations in a moderate amount of CPU time.
            
\begin{figure}
        \centering
        \digraph[scale=0.5]{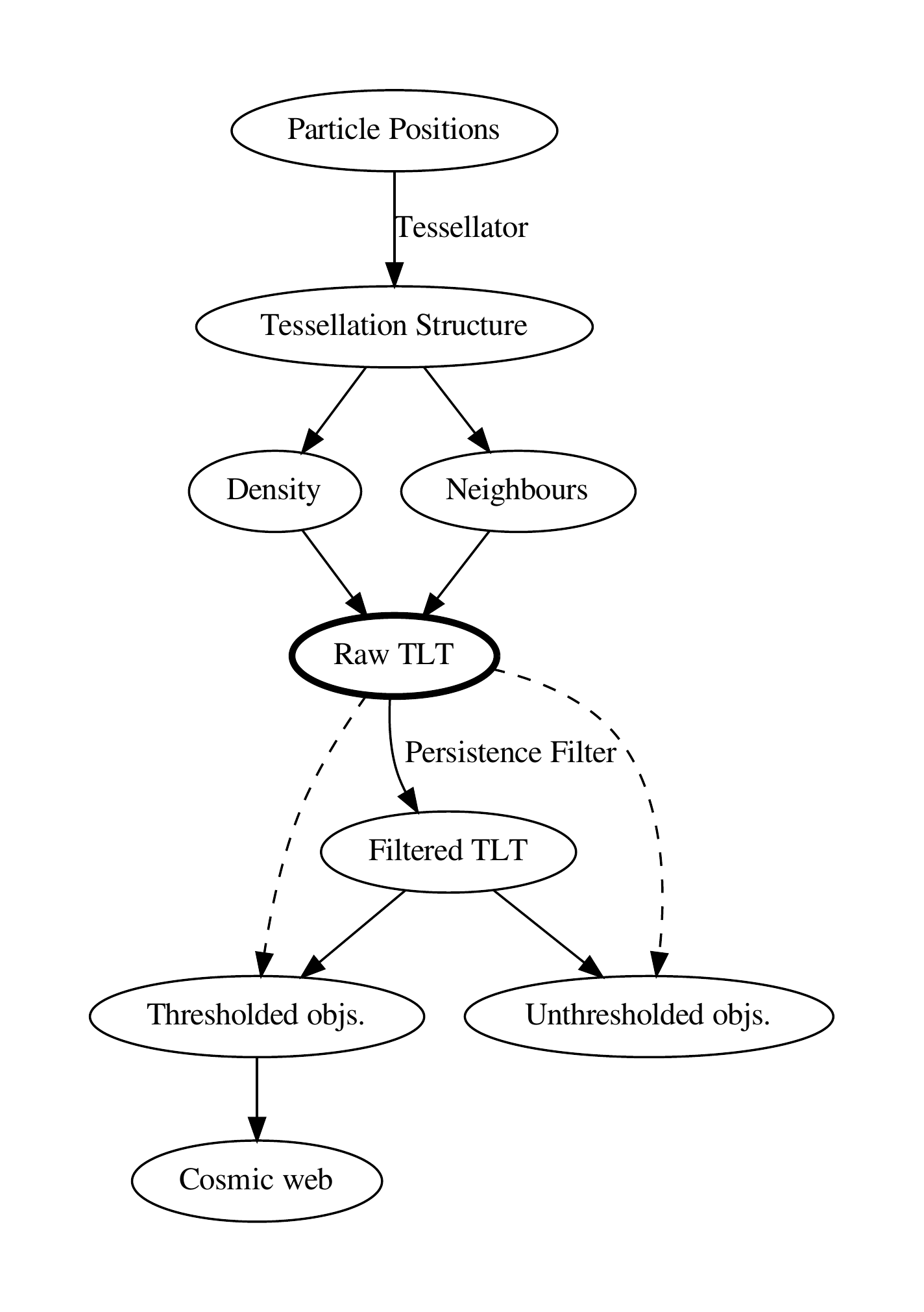}{rankdir=TB; 
        a [label="Particle Positions"];
        b [label="Tessellation Structure"];
        c [label="Density"];
        d [label="Neighbours"];
        e [label="Raw TLT", penwidth=3];
        f [label="Filtered TLT"];
        g [label="Thresholded objs."];
        h [label="Unthresholded objs."];
        i [label="Cosmic web"];
        a->b [label="Tessellator"];
        b->c;
        b->d;
        c->e;
        d->e;
        e->f [label="   Persistence Filter"];
        f->g;
        f->h;
        g->i;
        e->h [style="dashed"];
        e->g [style="dashed"];
        }
        \caption[Pipeline Overview]{Overview of the pipeline for the \methabb.}\label{fig:pipe_over}
      \end{figure}

      A schematic of the method is presented in \autoref{fig:pipe_over}. From the particle positions we obtain the tessellation structure (\autoref{sec:tessellation}). This structure provides us with a density estimate for each particle and a neighbourhood in the form of a list of neighbours from which we construct the hierarchical set of peaks (\autoref{sec:peak_const}).

    \subsubsection{Tessellation}\label{sec:tessellation}
     
     The formal basis of the \methabb is the unweighted Voronoi tessellation (VT) $\mathcal{T}$ in position space of the simulation particles $P$ whose positions act as its generators. The cells of the tessellation are the regions in this space to which a given generator is closest in Euclidean distance. This construction leaves us with cells in the shape of convex polytopes. As we are using periodic boundary conditions all these polytopes will be of finite extent.
     
     For this set of polytopes $P$ we use the volumes $V$ to define particle densities \begin{align}
      \rho_i = \frac{m_i}{V(p_i)}\text{, for }p_i\in P,\label{eqn:voro_dens}
     \end{align}
     and the shared faces to define a set of connections $E$. Particles are considered neighbours if their cells share a face.
     
     For the following we impose a strict density ordering on the full particle set of the simulation. While in practice it is very unlikely to find two particles with exactly the same attributed volume, this probability is not zero due to the finite computational precision. In case we do find two particles with the same density, we rank them randomly among themselves. We do not expect such degeneracies to affect any real applications as the available state space in units of the granularity imposed by numerical precision is simply too large, especially given that we use double precision floating point arithmetic.
    
    \subsubsection{Peak Tree Construction}\label{sec:peak_const}
    
    We traverse the full list of simulation particles in order of decreasing density starting from the highest density particle in the simulation. A single traversal is sufficient to construct the full \methabb. The procedure can also be reversed, starting from the least dense particle and proceeding in order of increasing density. This would produce a hierarchical tree linking all particles to local minima, similar to \textsc{ZOBOV} \citep{neyrinck_zobov_2008}. We do not pursue this farther in this paper.
    
    A peak $\pi_m = (i,\pi_n,k)$ is an object with a first particle $i\in\mathbb{N}^{N_{part}}_1$, a parent peak $\pi_n$ (initially $\pi_m$ itself) and a last particle $k$ (initially $i$). Both $m$ and $n$ are indices from the set $i\in\mathbb{N}^{N_{peak}}_1$, assigned in order of decreasing peak density. For each particle $j$ we keep two numbers, the peak $\pi_m$ it belongs to and $n_j$ the next particle (in order of decreasing density) in a link list connecting to that peak. One could also keep the rank of the previous particle in the peak's list to produce a doubly linked list that can be traversed in both directions.

    We scan all particles in the simulation once, from highest to lowest density rank. For each particle $j$ we examine the ranks of its neighbours. One of two cases then applies:
    \begin{enumerate}
     \item If the rank of the particle is higher than those of all its neighbours, we create a new peak object $\pi_m$ which represents a local maximum and is initiated with its first particle $i=j$.
     \item Otherwise, there are one or more neighbours with a higher rank. These particles will have been processed before the current one and will already have been assigned to a peak. We assign the current particle to the highest-ranked of these peaks. We also set it as the next particle $n_{j'}=j$ for all higher neighbours $j'$ which do not already have this link set to another particle. This leaves us again with two possible cases:
     \begin{enumerate}
     \item If all higher ranked neighbours belong to the same peak, the particle is assigned to this peak.
     \item If there are higher neighbours belonging to different peaks, the current particle represents a saddle between these peaks. (By virtue of the strict ordering, each peak has a highest density saddle that is processed first.) The particle $j$ is then assigned to the highest of these peaks, which is marked as the parent of all the lower ones. The particle sets for the lower peaks are now complete and they can be considered as subpeaks of the highest one.
    \end{enumerate}
    \item The last particle index of the peak to which particle  $j$ has been assigned is set to $j$.
    \end{enumerate}
    
    Each of the peak look-ups in this process is a actually a recursive operation that follows the chain of parent-subpeak links until it reaches a peak which is still considered as its own parent, i.e. is currently an independent peak. As the level surfaces start percolating, fewer and fewer independent peaks remain, until finally all peaks are (often indirect) children of the global maximum, the ultimate root of the \methabb .
    
\begin{figure}
      \centering
      \includegraphics[scale=0.75]{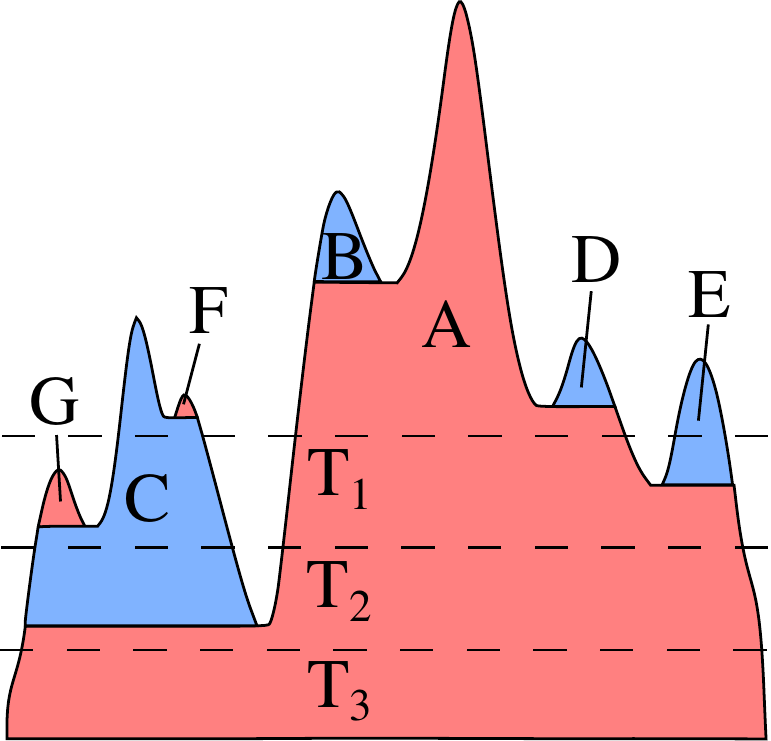}
      \caption{Schematic of the decomposition of a 1-D density distribution into peaks labelled alphabetically in decreasing peak density order and for three thresholds $T_{1}$ through $T_{3}$. The resulting tree structure is shown in \autoref{fig:peak_cartoon_tree}.}\label{fig:peak_cartoon}
    \end{figure}

\begin{figure}
      \centering
      \digraph[scale=0.5]{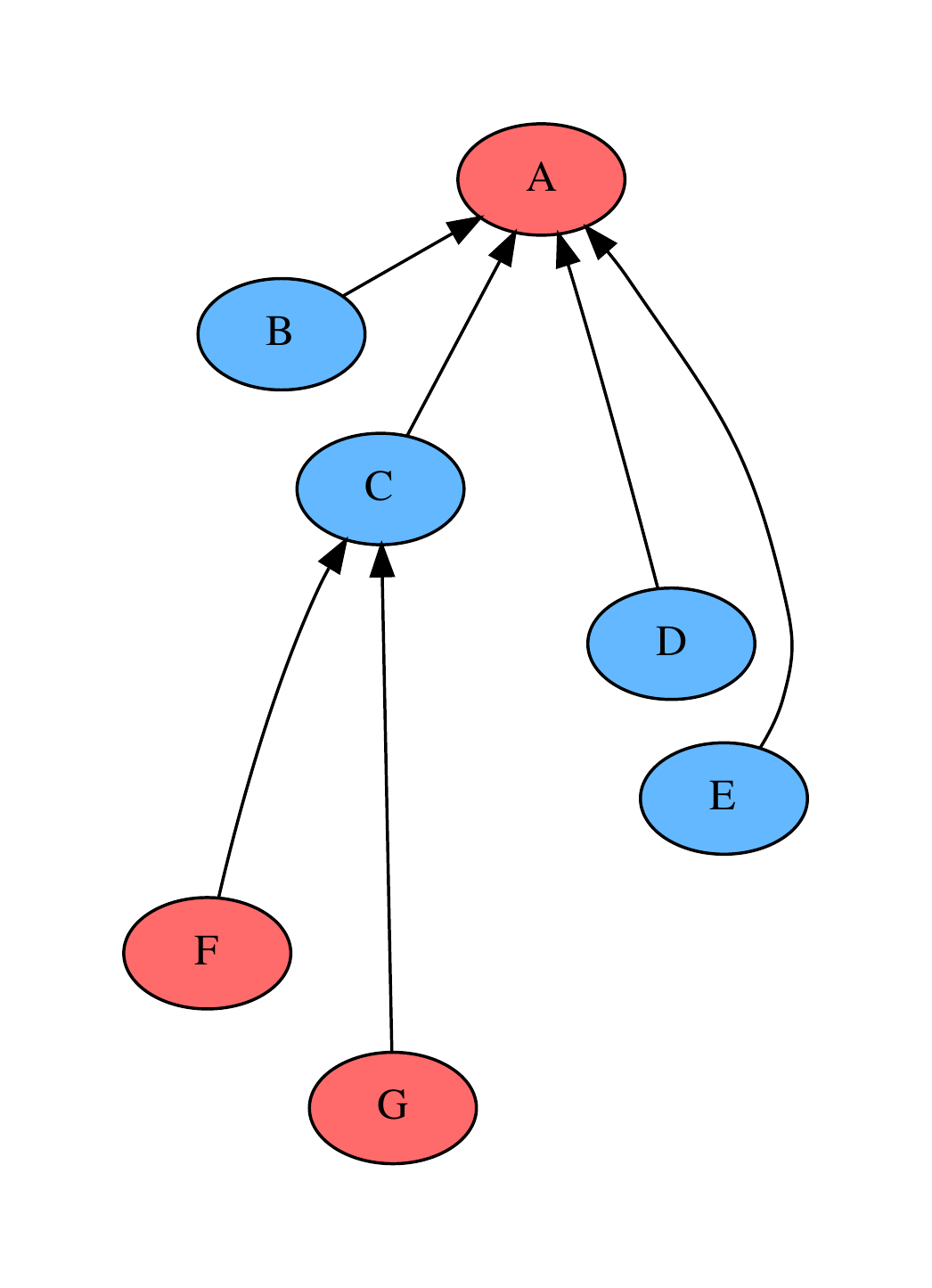}{

      rankdir=BT;
      ranksep=0.2

      A [fillcolor = indianred1, style = filled, pos = "0,0!"];
      F [fillcolor = indianred1, style = filled];
      G [fillcolor = indianred1, style = filled];
      B [fillcolor = steelblue1, style = filled];
      C [fillcolor = steelblue1, style = filled];
      D [fillcolor = steelblue1, style = filled];
      E [fillcolor = steelblue1, style = filled];

      invis1 [shape=none label="" width=.0];
      invis2 [shape=none label="" width=.0];
      invis3 [shape=none label="" width=.0];
      invis4 [shape=none label="" width=.0];
      invis5 [shape=none label="" width=.0];
      invis6 [shape=none label="" width=.0];
      invis7 [shape=none label="" width=.0];

      B->A;
      C->invis1->A [style=invis];
      C->A;
      D->invis2->invis1->A [style=invis];
      D->A;
      E->invis3->invis2->invis1->A [style=invis];
      E->A;
      F->invis6->invis5->C [style=invis];
      F->C;
      G->invis7->invis6->invis5->C [style=invis];
      G->C; 
      }
      \caption[Example Tree]{The structure of peak tree of the example in \autoref{fig:peak_cartoon} }\label{fig:peak_cartoon_tree}
    \end{figure}

    A 1D example of this hierarchical segmentation process is given in \autoref{fig:peak_cartoon}. The accompanying peak tree is given in \autoref{fig:peak_cartoon_tree}.

    Further physical properties of peaks are easily accumulated during construction of the \methabb, or can be calculated from it in post-processing. The simplest ones are the mass $M(\pi_m)$, volume $V(\pi_m)$ and the resulting mean density $\rho(\pi_m)$ associated with the peak $\pi_m$:
    \begin{align}
     M(\pi_m) &= \sum\limits_{j\in\pi_m} m_j \label{eqn:peak_mass} \\
     V(\pi_m) &= \sum\limits_{j\in\pi_m} V(p_j) \label{eqn:peak_vol} \\
     \rho(\pi_m) &= \frac{M(\pi_m)}{V(\pi_m)} \label{eqn:peak_dens}
    \end{align}
    
    Here the mass and volume of a peak exclude contributions from its subpeaks. If desired, these additional contributions can easily be included by using the tree structure to enumerate all subpeaks so that their mass and volume can be added to those of the main peak. This \methabb structure also allows for the very quick calculation of the properties of objects defined to be enclosed by any chosen density level surface. For each peak whose density range brackets the threshold, one simply follows the next-particle chain until the threshold is reached, and then, if required, adds the contribution of all subpeaks that join the main peak above this threshold. This will be used when we present various halo definitions in \ref{sec:halo_def}.

    Summarizing this section, we construct a decomposition of the set of particles in an N-body simulation into disjoint \emph{peaks}, each a set of particles. Each peak consists of all particles that are assigned to that peak and are reached from its \emph{peak particle} when traversing the linked list that leads away from it in descending density order. Each peak has a range in density from that of its peak particle down to (but not including) that of the first particle in the linked list which has an ascending path (and so is assigned) to a higher density peak, the \emph{saddle particle}\footnote{As we work with a discrete set of densities, defined only at particle positions, our saddle particle definition is based purely on its density rank and on those of its neighbours. When finding critical points we do not need to consider a density field, continuous, differentiable or otherwise, of the kind used by most other approaches. We also stress that our scheme considers only the unique \emph{highest} saddle point particle connecting a given peak to a higher one and ignores all others.}. The density of this saddle particle is a lower bound to densities in the peak and therefore sets its \emph{limiting density} $\rho_{lim}$. The peak that the saddle particle is assigned to becomes the parent of the peak under consideration. The hierarchical data structure created in this way can be used to find halo properties for many different halo definitions, as detailed in \ref{sec:halo_def}.
  
  \subsection{Persistence and the Choice of the Density Estimator}\label{sec:pers}
    
    The discretisation of a smooth density field by particles, as in an N-body simulation, can lead to problems when estimates for the underlying field are constructed from the particle positions. Such estimates typically show many low-amplitude peaks caused by sampling noise. The character of this noise is not immediately evident in our case, since in low-density, single-stream regions N-body particle distributions are typically subrandom (i.e. have lower variance than a Poisson process) because of the grid or glass distributions used for the initial particle load. At higher densities, however, and particularly inside halos, Poisson sampling is expected to be a good model and discreteness noise produces many weak local peaks. In this section we analyse discrete samples from spatially uniform or slowly varying Poisson processes to show how almost all such peaks can be eliminated by pruning the \methabb to retain only sufficiently \emph{persistent} peaks \citep[see][for closely related noise suppression procedures]{neyrinck_voboz_2005,neyrinck_zobov_2008}. The measure of persistence we use here was introduced by \cite{edelsbrunner_topological_2002} and further developed in \cite{zomorodian_computing_2005}. Persistence can be understood as a measure of the durability of a feature, in our case a peak, as we lower the density threshold. The more important a peak is and the stronger it overtowers its surroundings, the higher is its persistence. A related problem which we do not address in this paper is that on the boundaries between single- and multi-stream regions, caustics lead to sudden density jumps and the tessellation techniques we employ then produce significantly biased estimates of the local density \citep[see the detailed discussion in][particularly that around their Fig.13]{abel_tracing_2012}.

    We have tested two different estimators for the density at the particle positions, one based on the Voronoi tessellation (\autoref{eqn:voro_dens}), the other on its dual, the Delaunay tessellation. The Delaunay estimator, as introduced by  \cite{schaap_continuous_2000}, is similar to the Voronoi case already described, but takes the volume associated with each particle to be one  quarter of the sum of the volumes of its adjacent Delaunay tetrahedra:
    \begin{equation}
    \rho_{D} = \frac{m_P}{V_{D}} = m_P \left(\sum \limits_{c\in \mathcal{C}} \frac{1}{4}V(c)\right)^{-1},\label{eqn:del_rho}
    \end{equation}
    where $\mathcal{C}$ is the set of Delaunay tetrahedra which have a vertex on the given particle. Since each tetrahedron is spanned by four particles, this distributes the complete volume of the simulation over the particles, as in the Voronoi case.
    
    We define the persistence $r$ of a peak as the ratio of its peak density to its limiting density.  We then filter the \methabb by requiring all peaks to have persistence larger than some threshold. If a peak fails this criterion,  it is removed from the hierarchy and all its member particles are are inserted at the appropriate points in the density-ordered link list below the parent peak. Assuming that ``true" peaks are usually more persistent than ``noise" peaks, this can remove most of the latter while losing only the weakest of the former.
    Note that in order to allow consistent construction of objects containing only particles above a chosen density threshold, it is necessary to assign each particle in a filtered \methabb an additional variable, namely the density rank $k'$ of the saddle-point particle of its original peak in the unfiltered \methabb (see \autoref{sec:halo_def}).
    
\begin{figure}
    \centering
    \includegraphics{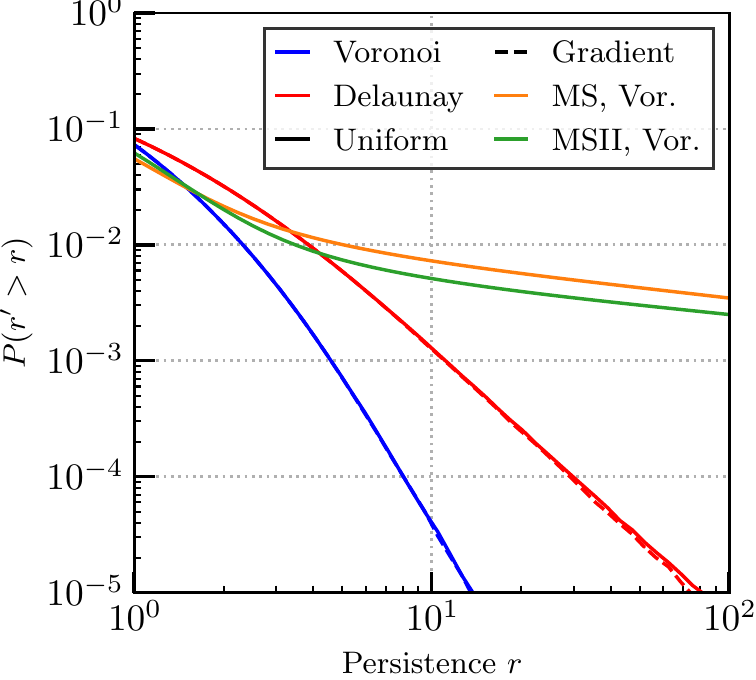}
    \caption[Poisson Persistence Results]{An application of the \method to Poisson-selected distributions of $10^7$ particles within a unit cube, assuming periodic boundary conditions, as well as to the two Millennium Simulations analysed later in this paper. The two test cases assume a uniform underlying density field and one with large-scale gradients and a single true peak. We show the probability $P(r'>r)$ that a random particle is identified as a peak with persistence exceeding the threshold $r$. The abundance of high persistence noise peaks is much lower for the Voronoi than for the Delaunay density estimator, but for both cases it is almost identical for uniform and for slowly varying underlying density distributions. The (Voronoi) peak distribution for the two Millennium Simulations is more than 99\% pure for $r>10$ }\label{fig:perc_test}
    \end{figure}

    To estimate the threshold needed to eliminate almost all noise peaks, we have studied the effect of filtering on the \methabb constructed for $10^7$ Poisson-selected points inside the unit cube when the Voronoi and Delaunay density estimators are used and periodic boundary conditions are assumed. We consider two cases for the underlying smooth density field, one where it is uniform and the other where it has the form $\rho\propto (1+0.3\sin{2\pi x})(1+0.3\sin{2\pi y})(1+0.3\sin{2\pi z})$. Since there is no true maximum in the first case and only one  in the second, essentially all peaks found by the \methabb can be considered as noise peaks in both cases. Before filtering, 7.319\% and 8.280\% of the particles are identified as peaks in the Voronoi and Delaunay cases, respectively, for the uniform underlying field. For the non-uniform field these numbers are 7.313\% and 8.283\%, respectively, showing that the abundance of noise peaks is insensitive to large-scale gradients in the underlying density.\footnote{Note that our version of the \methabb always uses neighbour lists derived from a Voronoi tesselation as described above.}  In \autoref{fig:perc_test} we show the fraction of particles which are identified as peaks after filtering out all peaks with persistence smaller than r. 
    
    We find that the Delaunay estimator results in a longer tail of noise peaks than the Voronoi estimator. For both, we find an asymptotically power law-like tail in the probability $P(r'>r)$ that the persistence $r'$ of a peak exceeds a given threshold value $r$. While this probability drops roughly as $r^{-2.2}$ for the Delaunay estimator, it drops as $r^{-4.6}$ for the Voronoi estimator. The behaviour is essentially identical for the cases with and without an underlying large-scale density gradient. Based on these results we will adopt the Voronoi density estimator for the rest of this paper and use a threshold of $r_{th} = 10$ when we wish to filter out noise peaks. This results in the elimination of 99.95\% of all noise peaks. at least in situations analogous to those of \autoref{fig:perc_test}.

\begin{figure}
    \centering
    \includegraphics{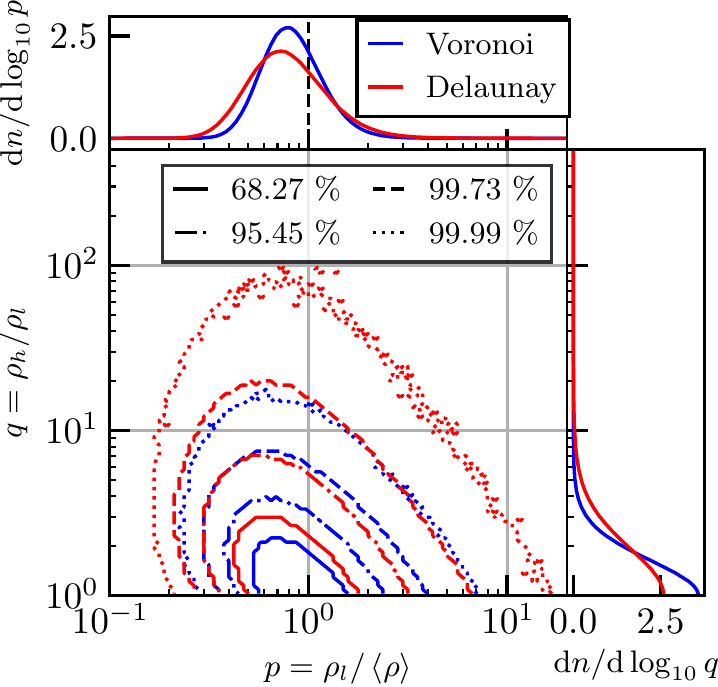}
    \caption[Pair Density Ratio]{The distribution of all pairs of Voronoi neighbours in the uniform density test of \autoref{fig:perc_test} in the plane of density ratio $q$ against the normalised value $p$ of the smaller of the two densities.  Blue and red lines show equidensity contours of these distributions in the Voronoi and Delaunay cases, respectively. The four different line styles indicate contours enclosing \{68.27,95.45,99.73,99.99\} percent  of the pairs. The panels on the $x$- and $y$-axes show the marginalized distributions in each variable. The distributions are broader and the tails are longer for both variables in the Delaunay case.}\label{fig:pair_rat}
    \end{figure}

    The difference in behaviour between the Delaunay and Voronoi
    estimators can be clarified if one takes a look at the distribution of absolute and relative densities for pairs of neighbours. A 2D scatter plot of this distribution is given for our uniform Poisson test case, together with its 1D marginal distributions in \autoref{fig:pair_rat}. We find that the
    marginal distributions of both variables are noticeably broader
    and have substantially longer tails in the Delaunay case.  In both cases there is only a weak correlation between the variables.
    This result might seem surprising, since the volume used in the Delaunay estimator is, on average, four times larger than that used in the Voronoi estimator, suggesting that the variance induced by the underlying Poisson point distribution might be four times smaller. Wald's equation and the Blackwell-Girshick equation \citep{blackwell_functions_1946} give us the relative standard deviation $\sigma(V_D)/\left<V_D\right>$ of the sum of $N=27.1\pm6.7$ \citep{okabe_spatial_2000} hypothetically independent and identically distributed (iid) tetrahedron volumes using the known normalised single-cell standard deviation $\sigma(V_{DT})/\left<V_{DT}\right> \approx 0.83$: 
    \begin{align}
    \begin{split}
      \frac{\sigma(V_D)}{\left<V_D\right>} &= \sqrt{\frac{1}{\left<N\right>}\left(\frac{\sigma(V_{DT})}{\left<V_{DT}\right>}\right)^2+ \left(\frac{\sigma(N)}{\left<N\right>}\right)^2}\\
      &\approx 0.29\label{eqn:iid_del}
    \end{split}
    \end{align}
    
    This argumentation is overly naive, however. Both the Voronoi cell volume and the summed Delaunay volume of \autoref{eqn:del_rho} are determined entirely by the positions of the \emph{same} set of particles, the Voronoi neighbours used by the \methabb. It turns out that the relative standard deviation of the Voronoi cell volume $\sigma(V_V)/\left<V_V\right> \approx 0.42$ is smaller than that of the summed Delaunay volume $\sigma(V_D)/\left<V_D\right> \approx 0.56$, almost twice as large as for the iid-case considered in \autoref{eqn:iid_del}. In addition, the distribution of  $V_D$ has longer tails than that of $V_V$, analogous to the result for peaks shown in \autoref{fig:perc_test}. \footnote{These results are consistent with the conclusion of \cite{pandey_exploring_2013} that the Voronoi and Delaunay estimators have similar variance, because these authors calculated the variance in the (interpolated) density estimate at a random point in space, whereas we calculate it at the position of a random particle.}
    
    For the analysis of the Millennium Simulations which we carry out later in this paper, it is of interest to know what fraction of the final filtered peak catalogue may still be spurious. (The complement of this quantity is often referred to as the catalogue purity.) In \autoref{fig:perc_test} we thus show the fraction of all particles in these simulations which are identified as peaks with persistence exceeding $r$. As expected, this declines much more slowly than in our noise-dominated test simulations, roughly as $r^{0.3}$ over the range $10<r<100$. Interestingly, the total peak fraction (i.e. for $r>0$) is slightly lower in the Millennium Simulations than in our tests, suggesting that the strong gradients associated with the true clustering have somewhat suppressed the noise peak abundance. At large $r$, the peak fraction in the MSII is about 30\% lower than in the MS, reflecting the smaller fraction of the total mass in peaks with individual mass close to the resolution limit. For $r>10$, the filter we adopt below, the fraction of particles which are peaks is more than two orders of magnitude smaller in the Voronoi test simulations than in either Millennium Simulation, suggesting that their filtered peak catalogues are more than 99\% pure.  
    
    Note that this filtering approach will necessarily lead to the exclusion of some low amplitude but legitimate peaks, in addition to the noise peaks we are trying to eliminate. We investigate and quantify this issue further in \autoref{sec:abund_unth} using our cosmological simulation data. The only way to mitigate this problem is to choose a moderate persistence filter and to focus on issues which depend primarily on high-contrast peaks.  We consider $r_{th} = 10$ as an acceptable compromise for most purposes, but will use the unfiltered \methabb for some analyses where filtering introduces problems. 

  \subsection{Halo Definitions}\label{sec:halo_def}
  
    The \methabb enables halo catalogues to be constructed
    according to a number of halo definitions, depending on the choices made for various parameters. The first is the persistence threshold $r_{th}$, which we have just discussed. Once this is chosen, every peak particle in the filtered \methabb corresponds to an ``unthresholded" object similar to a \verb+SUBFIND+ subhalo in that it contains all particles associated with the peak, and all of these have $\rho > \rho_{lim}$, where $\rho_{lim}$ is the density at the highest saddle point connecting to a higher peak and varies from object to object. Alternatively, one can choose a density threshold and consider haloes to consist only of higher density particles. In this case, only objects with peak density above threshold are retained, those with  $\rho_{lim}$ below threshold corresponding to main haloes and the rest to subhaloes. Finally for both unthresholded and thresholded haloes (denoted UH and TH below) one can either include or exclude subhaloes when enumerating the particle content (denoted UH$^+$, UH$^-$ and TH$^+$, TH$^-$ below). As we will see, a definition thresholded at about 80 times the mean density and including substructures (TH$^+$(80)) produces haloes very similar to the standard FoF algorithm.  

    \subsubsection{Unthresholded Haloes}
    
    Each UH\scp is made up of all particles enclosed by a level surface at its limiting density, $\rho_{lim}$. Its mass and volume are thus the sums of the masses and volumes which the \methabb assigns to the main peak and all its subpeaks. In contrast, the particle content, mass and volume of a UH$^-$ 
    are identical to those assigned to its peak particle by the \methabb. The tree in \autoref{fig:peak_cartoon_tree} contains seven UHs in either definition. Each UH\scp contains all peaks below it in the tree, so that G is a UH on its own (as it is lacking substructure, UH\scp and UH\scm are identical in this case) but is also included in the UH\scp's C and A. The UH\scm's, on the other hand, are a disjoint partition of the mass, and so correspond to the seven different coloured areas in the figure.
    
    A UH can span a density range lying anywhere within the full density range of the simulation provided the ratio of its maximum and minimum densities exceeds the adopted persistence threshold. Indeed, the UH associated with the highest peak always spans the full density range of the simulation. When compared to the objects identified by a standard halo/subhalo finder, an individual UH may correspond to a subhalo, to a main halo together with some surrounding material, or may not appear at all if its peak density lies below the density threshold bounding the ``classical" haloes.

    \subsubsection{Thresholded Haloes}

    For any given threshold density, every peak with peak particle density above threshold and limiting density below threshold corresponds to a \emph{thresholded halo} (TH). In the TH$^-$ case this halo consists of all particles in the link list below the peak particle which have density exceeding the threshold. In the TH\scp case, it additionally contains all particles in subpeaks that join the main peak (i.e. have saddle-point density $\rho_{lim}$) above threshold. For filtered \methabb's, particles which have been added to the link list as a result of suppression of a low-persistence peak, should only be included in the TH if the limiting density of their original (pre-filtering) peak is above threshold. This definition establishes TH\scp's as the material enclosed by isodensity bounding surfaces which follow the faces separating pairs of Voronoi cells with densities on opposite sides of the threshold.
    
    As an example, the 1D density structure in \autoref{fig:peak_cartoon} would give the thresholded peak sets (and therefore TH\scm sets) \{A,C,E\}, \{A,C\} and \{A\} for the thresholds T\textsubscript{1}, T\textsubscript{2} and T\textsubscript{3}, respectively. If we include the substructures to obtain the TH\scp definitions the sets would be \{A+B+D,C+F,E\}, \{A+B+D+E,C+F+G\} and \{A+B+C+D+E+F+G\}. In each of these cases only the part above threshold would be included for peaks which extend to lower densities. If the \methabb were filtered so that only the largest peaks, A and C, pass the persistence threshold, then all particles from E would be included in the TH\scp for A at thresholds $T_2$ and $T_3$ but none of them at $T_1$. All particles from B and D would be included in A, and all particles from F included in C for all thresholds. All particles from G would included in C for the lower thresholds but not for $T_1$.
    
    For thresholded haloes, the formulae for peak mass (\autoref{eqn:peak_mass}) and volume (\autoref{eqn:peak_vol}) can be used as before, provided the sums are extended only over particles with densities above \rt, and for which, in the case of particles added to the link lists as a result of persistence filtering, the limiting density of the original (pre-filtering) peak is also above threshold. The mean density in \autoref{eqn:peak_dens} should then also use these modified quantities.
    
    The FoF algorithm also defines haloes to be connected objects above a given bounding density which it estimates crudely using the nearest neighbour distance. As a result it, produces haloes quite similar to our TH\scp's for suitably matched FoF linking length and \methabb density threshold. The differences in density estimate and neighbour definition mean that the correspondance is far from exact, however. We describe this in more detail in \ref{sec:massf}. A closely related discussion can be found in \cite{more_overdensity_2011}.
 
 \subsection{Derivative Quantities}\label{sec:deriv_quant}
  
    The structure of the \method and the information it contains about the density at the position of each simulation particle allow us to characterise objects selected according to any of the above definitions in a number of new ways, in addition to providing simple summary quantities such as mass, volume and mean density. We discuss some of these additional properties in this paper in order to illustrate the potential of the \methabb; they will be explored in more detail in forthcoming publications.
    
    
    The characterisation of substructure in haloes is easily accomplished, since child peaks of the main halo peak are subhaloes defined in a very similar way to those identified, for example, by the \textsc{subfind} algorithm; the main difference is that here there is no attempt to require them to be gravitationally self-bound. Properties such as subhalo mass, position, velocity, local environment density and internal structure are all easily obtained from the \methabb
    in conjunction with the corresponding simulation snapshot.
    
    When using the \methabb to construct density profiles for dark matter haloes, it is easy to include or exclude the mass in substructures, just as for substructure finders such as \textsc{Subfind} or \textsc{rockstar}. As an alternative to standard profiles which average the density in spherical or elliptical shells and quote mean density as a function of radius, the \methabb enables construction of profiles of the form $M(>\rho)$ giving the total mass of halo particles for which the \emph{local} density exceeds $\rho$. This formulation involves no smoothing of the density field other than that imposed by particle discreteness, and so does not smooth away the high densities in substructures. In addition it makes no assumption about halo shape or internal structure. This is advantageous for some purposes. For example, it allows a much improved estimate of the total dark matter annihilation rate within a halo, since this is proportional to the sum over all halo particles of the product of the particle mass and \emph{local} dark matter density.
    
    
    We can also define new measures of halo concentration based on this alternative approach to density profiles. A simple one involves the ratio of the masses above two predefined density thresholds. The larger the mass fraction above the higher threshold, the more concentrated is the halo. This approach has the advantages that it can easily include or exclude the subhaloes\footnote{As before, in the TH cases it is important to ensure that if the \methabb is filtered, then  particles from suppressed subpeaks with saddle-point density below the threshold are excluded when scanning the link lists.} and it avoids the often implicit assumption that the halo mass distribution is approximately symmetric or can be fit acceptably by a parametric form such as the NFW profile.
    For many haloes this parametric form is, of course, a good fit. The new measure can be calibrated to return the standard value of concentration for such haloes by using the NFW formula to calculate the ratio of the masses above the two predefined thresholds and then inverting the obtained mass-ratio -- concentration relation.

    
    Sizes, shapes, mean velocities and spins are also easily estimated for all our haloes, including or excluding subhaloes and for any threshold density exceeding their limiting density. Standard methods from the literature can be applied to particle lists obtained from the \methabb together with positions and velocities taken from the corresponding simulation snapshot. For example, halo shapes and orientations can be obtained as a function of density threshold using the reduced tensor of inertia method of \cite{allgood_shape_2006}.
    The following subsections give more detail for some of the properties which are investigated later in this paper using the Millennium simulations. 
    
    \subsubsection{Density Profiles}\label{sec:den_prof}
    
    Traditionally, the density profile of a simulated object is given as the mean mass density in spherical shells surrounding an appropriately chosen centre, either including all particles, or excluding those that are unbound or are identified as belonging to a subhalo. Thus the profile is a one-dimensional function of the form $\rho(r)$ . While this definition is simple and convenient, it ignores the fact that simulated haloes are far from spherical and contain substantial substructure. It is straightforward to address the first issue, at least partially, by averaging over ellipsoidal rather than spherical shells, but the smoothing of substructure caused by any such averaging can be problematic, most notably when estimating collision or annihilation rates for dark matter particles. 
    
    Using the \methabb it is easy to construct an alternative one-dimensional profile which makes no assumption about the symmetry properties of the halo and involves no additional smoothing. Since we have a density estimate for every halo particle, we can, for any of our halo definitions, simply construct $M(\rho)$, the total mass in halo particles with individual densities exceeding $\rho$. As before, when using filtered catalogues it is important to include particles from suppressed ``noise" peaks only if the corresponding saddle-point density is above threshold. For halo definitions which include subhalo particles (i.e. UH$^+$ and TH$^+$) two different schemes suggest themselves.
    
    In the first scheme, particles belonging to a subhalo contribute to $M(\rho)$ only for $\rho < \rho_{lim}$.
    The mass as a function of density then jumps by the total
    mass of the subhalo as $\rho$ crosses its saddle-point density in exactly the same way as the mass enclosed within radius $r$ jumps when $r$ crosses the distance of the subhalo from the centre of its parent. Because of this analogy, we refer to the mass-density profile defined in this way as a \emph{pseudo-radial profile} (PRP) $M_{pr}(\rho)$. For a peak $\pi_m$ we then have
    \begin{align}
      M_{pr}(\pi_m,\rho) = \sum\limits_{j\in P_m(\rho)}m_j + \sum\limits_{\pi_k\in C_m(\rho)} M(\pi_k), \label{eqn:pr_prof}
    \end{align}
    where $P_m(\rho)$ is the set of all particles directly connected to $\pi_m$ at densities above $\rho$, $m_j$ is the mass of particle $j$, and $C(m,\rho)$ is the set of all children $\pi_k$ of $\pi_m$ for which $\rho_{lim,k}\geq\rho$. The subhalo masses $M(\pi_k)$ are defined as in \eqref{eqn:peak_mass}. This profile definition applies both to unthresholded  and to thresholded objects. Here and in the following, the possible densities are, of course, limited by $\rho\geq\rho_{lim,m}$ for unthresholded and by $\rho\geq\rho_{th}$ for thresholded objects. 
    
    The second scheme includes each subhalo particle in $M(\rho)$ for all $\rho$ smaller than its own individual density. With this definition $M(\rho)$  is the total mass within the halo boundary for which the local density exceeds $\rho$, regardless of whether the mass is part of the main halo, of a subhalo, or of a suppressed noise peak. We therefore refer to such a profile constructed for peak $\pi_m$ as its \emph{total-mass profile} (TMP) $M_{tot}(\pi_m,\rho)$. It can be expressed as:
    \begin{align}
     M_{tot}(\pi_m,\rho) = \sum\limits_{j\in P_m(\rho)}m_j + \sum\limits_{\pi_k\in C_{S,m}}\sum\limits_{l\in P_k(\rho)}m_l\label{eqn:tot_prof}
    \end{align}
    where the same definitions as in \eqref{eqn:pr_prof} apply, and, in addition, $C_{S,m}$ is the set of all subpeaks under $\pi_m$. In the case of thresholded objects (with threshold $\rho_{th}$) this last definition has to be altered, as only the subtree set $C_{S,j}(\rho_{th})$ of children with $\rho_{lim}>\rho_{th}$ should be included in the second term. This profile provides the information needed to estimate the total dark matter collision or annihilation rates, since these are proportional to the integral of $\rho M_{tot}(\rho)$ over the full density range of the halo. 
    
    If one refrains from including the mass in subhaloes we obtain the \emph{main-halo profile} (MHP):
    \begin{align}
     M_{mh}(\pi_m, \rho) = \sum\limits_{j\in P_m(\rho)}m_j,\label{eqn:cent_prof}
    \end{align}
    where the notation is as in \eqref{eqn:pr_prof}. This is also equivalently defined for unthresholded and thresholded objects (i.e. UH$^-$ and TH$^-$).
    
    Lastly one can define a mass-density profile  specifically for the mass in subhaloes, the \emph{subhalo profile} (SHP) $M_{sub}(\pi_m, \rho)$. The SHP is simply the difference of the TMP and the MHP:
    \begin{align}
     M_{sub}(\pi_m,\rho) &= M_{tot}(\pi_m,\rho) - M_{mh}(\pi_m,\rho)\\
     &= \sum\limits_{\pi_k\in C_{S,j}}\sum\limits_{l\in P_k(\rho)}m_l,\label{eqn:sub_prof}
    \end{align}
    where the notation is the same as in \eqref{eqn:tot_prof}.
    
    Given any fitting formula for the spherically averaged radial density profile, $\rho(r)$, a corresponding mass-density profile of the kind we have been discussing
    is obtained easily in parametric form if an analytic integral for the enclosed mass is available. For example, the NFW profile of \cite{navarro_universal_1997}  is normally written,
    \begin{align}
    \rho(r) = \frac{\delta~\rho_{crit}~r_s^3}{r~(r_s + r)^2},
    \label{eqn:NFW}
    \end{align}
    where $\rho_{crit}$ is the critical density, and $\delta$ and $r_s$ are characteristic scales for halo overdensity and radius, respectively. The mass-density profile is then simply
    \begin{align}
    \rho(s) &= \delta~\rho_{crit}~s^{-1}(1 + s)^{-2},\notag \\ M(s) &= 4\pi\delta~\rho_{crit}~r_s^3~\left[\ln{(1+s)} - \frac{s}{1+s}\right],~~0 < s < \infty ,
    \label{eqn:NFW2}
    \end{align}
    with $s=r/r_s$. A direct inversion to eliminate $s$ from these equations and obtain $M(\rho)$ or $\rho(M)$ is possible, but yields a rather complex result which we give in Appendix A.
    
  \subsubsection{Substructure Fraction}
  
    With the profiles introduced in last section we can, for any peak $\pi_m$, measure the mass fraction in substructure above any density $\rho$ that lies above $\rho_{lim, m}$ and below the density  of the peak particle. The ratio of the SHP to the TMP gives the fraction of all mass at such densities which is part of a subhalo,
    \begin{align}
     f_{sub}(\pi_m, \rho) = \frac{M_{sub}(\pi_m, \rho)}{M_{tot}(\pi_m, \rho)}\label{eqn:sub_frac}
    \end{align}
    One has to be careful when comparing this substructure fraction with one defined in the traditional way using radial density profiles. The latter is better approximated by
    \begin{align}
     f'_{sub}(\pi_m, \rho) = \frac{M_{pr}(\pi_m, \rho) - M_{mh}(\pi_m, \rho)}{M_{pr}(\pi_m, \rho)}.
    \end{align}
    With this definition the mass of each subhalo is counted all at once when it joins the main object, i.e. when $\rho$ drops to the saddle-point density at which subhalo and halo are linked. This is similar to the way in which traditional definitions include the mass of a subhalo in the total enclosed mass of its parent only for radii exceeding the distance of the subhalo from halo centre.
    
    When estimating the total subhalo mass fraction for a halo these two definitions. For unthresholded haloes $f_{sub}(\pi_m, \rho_{lim}) = f'_{sub}(\pi_m, \rho_{lim})$ and for thresholded haloes $f_{sub}(\pi_m, \rho_{thresh}) = f'_{sub}(\pi_m, \rho_{thresh})$.
  
  \subsubsection{Concentration Definitions}\label{sec:mass_conc_def}

    In a very general sense the concentration of a dark matter halo characterises the relative amount of matter residing at high densities. Mass-density profiles of the kind introduced above thus lead naturally to a variety of concentration definitions. A particularly simple example
    takes the fraction of halo mass which lies above some
    characteristic high density, or, more generally, the ratio
    $M(\rho_1)/M(\rho_2)$ of the profile values for two different densities, $\rho_1 > \rho_2$. Such a ratio ensures that the concentration measure does not depend on overall mass scale, and that concentration increases with the fraction of halo mass at large density. Thus we can define,
    \begin{align}
     c_{tot}(\pi_m,\rho_1,\rho_2) = \frac{M_{tot}(\pi_m,\rho_1)}{M_{tot}(\pi_m,\rho_2)},
    \end{align}
    with $\rho_1>\rho_2$ and $M_{tot}$ as defined in \eqref{eqn:tot_prof}. This definition may be particularly appropriate for estimating the DM annihilation signal as it measures the mass at high density in all regions, not just that near the centre of the main halo.
    
    For comparison with standard concentration measures based on spherically averaged radial density profiles, a definition based on our pseudo-radial profiles may be more appropriate,
    \begin{align}
     c_{rp}(\pi_m,\rho_1,\rho_2) = \frac{M_{trp}(\pi_m,\rho_1)}{M_{rp}(\pi_m,\rho_2)},
    \end{align}
    where $\rho_2$ would then be taken to be close to the bounding density of a conventionally defined halo, and $\rho_1$ would be taken to be 10 or 100 times higher.
    
    A conversion between concentrations defined in this way and those conventionally obtained from fits to spherically averaged radial density profiles can easily be obtained by integrating the fitting functions over radius and taking the ratios of the enclosed masses within the radii where the density crosses the adopted values of $\rho_1$ and $\rho_2$. We discuss this further for the particular case of NFW concentrations in Appendix A.

 \subsection{Halo Bias}\label{sec:bias_def}
    
    In studies of cosmological structure, the bias parameter $b$ is conventionally defined as the ratio of the clustering amplitude on large scale of some set of tracer objects (e.g. galaxies, haloes or galaxy clusters) to that of the dark matter. In this paper, we will follow
    \cite{gao_assembly_2007} and estimate $b$ as the ratio of the halo-dark matter cross-correlation to the dark matter autocorrelation. This procedure has the advantage that the very large number of dark matter particles available in the simulations we use leads to relatively precise bias measurements, even for quite small samples of haloes.  
  
    Specifically, we slightly depart from \cite{gao_assembly_2007} and estimate $b$ as the value that minimises
    \begin{equation}
      \sum \limits_{i=1}^4 \left(\frac{\xi_{hm,i}}{\xi_{mm,i}}-b\right)^2,
    \end{equation}
    where $\xi_{hm,i}$ and $\xi_{mm,i}$ are halo-dark matter cross-correlation and dark matter autocorrelation estimates in bin $i$, and our four bins are spherical shells of equal logarithmic width spanning the separation range $6< r/\left(\si{\Mpch}\right) < 20$. Unlike \cite{gao_assembly_2007} we do not take logarithms of the correlations in the numerator and denominator because, unlike them, we will also have to deal with negative values. We calculate cross- and auto-correlations on a cubic grid with $512^3$ cells using nearest grid point deposition. The grid thus has cell size $\SI{0.977}{\Mpch}$ in the larger of our two simulations and is five times smaller in the other.  This is sufficient to get precise correlation estimates on the scales where we need them.

  \section{The Simulations}
  
    We have applied the \method analysis methods outlined in the previous section to the Millennium \citep[MS][]{springel_simulating_2005} and Millennium II \citep[MSII][]{boylan-kolchin_resolving_2009} simulations. These two $10^{10}$ particle dark-matter-only simulations were carried out assuming a flat $\mathrm{\Lambda CDM}$ cosmology with the parameters given in \autoref{tab:sim_params}. These were chosen to match the first-year results from the \textsc{WMAP} satellite \citep{spergel_first-year_2003}. Although these are not formally consistent with more recent 
    measurements \citep[e.g.][]{planck_planck_2018, abbott_dark_2018}, the shifts are too small to be significant for our purposes. \autoref{tab:sim_params} also gives some of the numerical parameters defining the simulations, in particular, the total particle number, $N_{part}$, the mass of an individual particle, $m_{part}$, the side-length of each simulation's periodic cubic volume, and its Plummer-equivalent gravitational softening, $\epsilon$.
    
    The two simulations differ in length resolution by a factor of 5 and in mass resolution by a factor of 125. The larger box provides better statistics on large-scale structure, while the smaller one allows us to check  how the quantities we derive are affected by spatial resolution and discreteness noise. As we will show below, there is generally good agreement between the two simulations once noise peaks are filtered out as discussed in \autoref{sec:pers}.

    \begin{table}
      \caption{Parameters of the two simulations used in this work: the Millennium (MS) and the Millennium II (MSII).}\label{tab:sim_params}
      \centering
      \begin{tabular}{lcc}
      
	\hline
	& MS & MSII\\
	\hline
	$\Omega_{dm}$ & \multicolumn{2}{c}{0.205}\\
	$\Omega_{b}$ & \multicolumn{2}{c}{0.045}\\
	$\Omega_{\Lambda}$ & \multicolumn{2}{c}{0.75}\\
	$h$ & \multicolumn{2}{c}{0.73}\\
	$\sigma_{8}$ & \multicolumn{2}{c}{0.9}\\
	$n_s$ & \multicolumn{2}{c}{1}\\
	\hline \\
	$N_{part}$ & \multicolumn{2}{c}{$2160^3$}\\
	$m_{part}/\left(\si{\Msolh}\right)$ & $8.61\cdot 10^{8}$ & $6.88\cdot 10^{6}$ \\
	$L_{box}/\left(\si{\Mpch}\right)$ & $500$ & $100$ \\
	$\epsilon/\left(\si{\kpch}\right)$ & $5$ & $1$ \\
	\hline
      \end{tabular}
    \end{table}

\section{Thresholded cosmic web}\label{sec:above}

In this section we use the \method to study the large-scale structure of equidensity surfaces of the cosmic mass density field. While others have used this same Voronoi density estimate to study the one-point density distribution within the Millennium simulations \citep{pandey_exploring_2013,stucker_median_2018}, we here explore examples of how the \methabb enables quantitative study of many geometrical and topological aspects of cosmic structure.  In this and later sections we consider the simulated density distributions only at $z=0$, but it will clearly be worthwhile also to explore how these aspects of the distribution evolve with time.

\subsection{Total Mass and Volume}\label{sec:mv_above}

We begin our discussion of the thresholded density field by considering the total mass and volume which lie above threshold as a function of \rt . Both these quantities can be derived from the one-point density distribution and are independent of the geometry of equidensity surfaces. They were already considered in considerable detail by \cite{pandey_exploring_2013}. However, they can also be thought of as the sum of the masses of all thresholded haloes (TH$^+$) as a function of \rt.\footnote{Note that this equivalence holds strictly only if the \methabb is not persistence filtered, although differences are very small for the persistence thresholds we use in this paper.} We recap the results here because they are useful for comparison with the properties of individual large objects which we discuss in the next section.  In \autoref{fig:ms1_perc} and \autoref{fig:ms2_perc} blue and orange curves show the fractions of the total mass $\sum M$ and the total volume $\sum V$ above threshold as a function of \rt for the MS and the MSII respectively.

The total mass and volume both increase smoothly with \rt in both simulations, and the differences between them are quite small. The mass fraction is already close to unity at $\rt = 100 \rmean$, with values of 45\% and 55\% in the MS and the MSII, respectively. By the time the threshold has dropped to the mean, these mass fractions have risen to 72\% and 88\%. The difference between the two simulations is a direct consequence of the difference in their resolution. The MSII resolves haloes down to a mass limit which is two orders of magnitude lower than in the MS. The material of these low-mass objects is considered to be at low density in the MS but at high density in the MSII. Thus, at every \rt the high-density mass fraction in the MSII exceeds that in the MS. At extremely high resolution, the high-density mass fraction in a $\mathrm{\Lambda CDM}$ cosmology is expected to be above 90\% even for $\rt = 100 \rmean$ \citep{angulo_birth_2010}. 

The total volume above \rt evolves roughly as $\sum V \propto \rt^{-1}$, as is to be expected given the slow change with threshold of  $\sum M$. The slope of $\sum V$ is slightly shallower for the MSII. This is again related to its higher resolution. At low densities (i.e.  $\rho \sim \rmean$) the boundaries of objects are broadened in the MS by its lower resolution, and this somewhat increases the volume assigned to them. In both simulations we find that just below 0.1\% of the total volume has density exceeding $100\rmean$, while $\sim 8\%$ and $\sim 6\%$ of the total volume has density exceeding the mean in the MS and the MSII, respectively.

\subsection{Percolation}\label{sec:perc}

\begin{figure}
 \includegraphics[width=\linewidth]{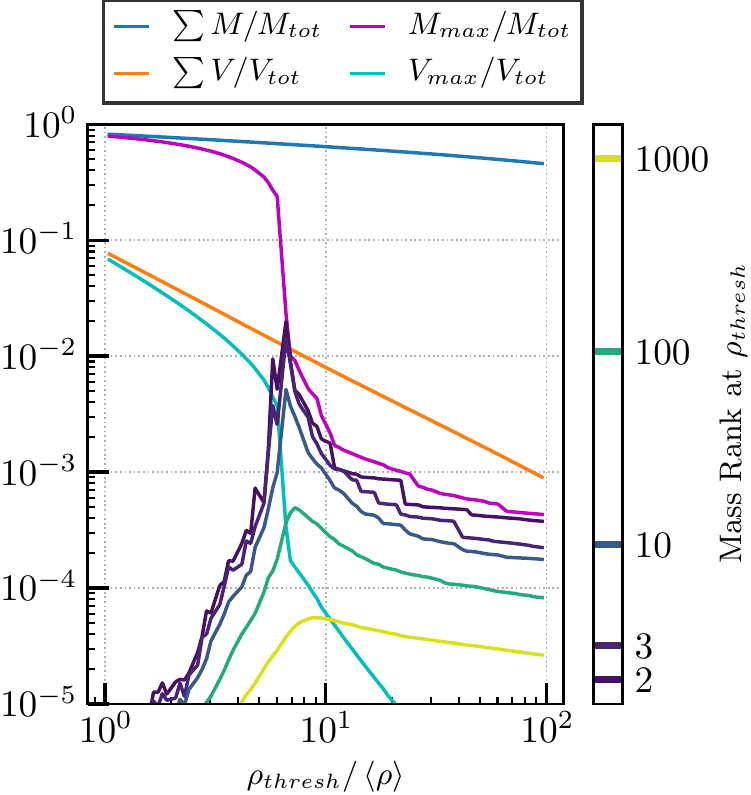}
 \caption{Total mass and volume above a given threshold $\rho_{thresh}$ in the 
  MS ($\sum M$ and $\sum V$) and also the mass and volume in the most massive 
  object ($M_{max}$ and $V_{max}$). Masses are, in addition, shown as a function 
  of \rt for the second, third, tenth, hundredth and thousandth most massive 
  objects (as identified on the colour bar). While the global quantities vary 
  smoothly, a clear phase transition is visible in the other curves over the 
  range $6 \lesssim \rt/\rmean \lesssim 7$ as many of the largest objects join 
  together to form a single percolating structure.  This structure can be 
  considered as defining the cosmic web.}\label{fig:ms1_perc} 
\end{figure}

\begin{figure}
 \includegraphics[width=\linewidth]{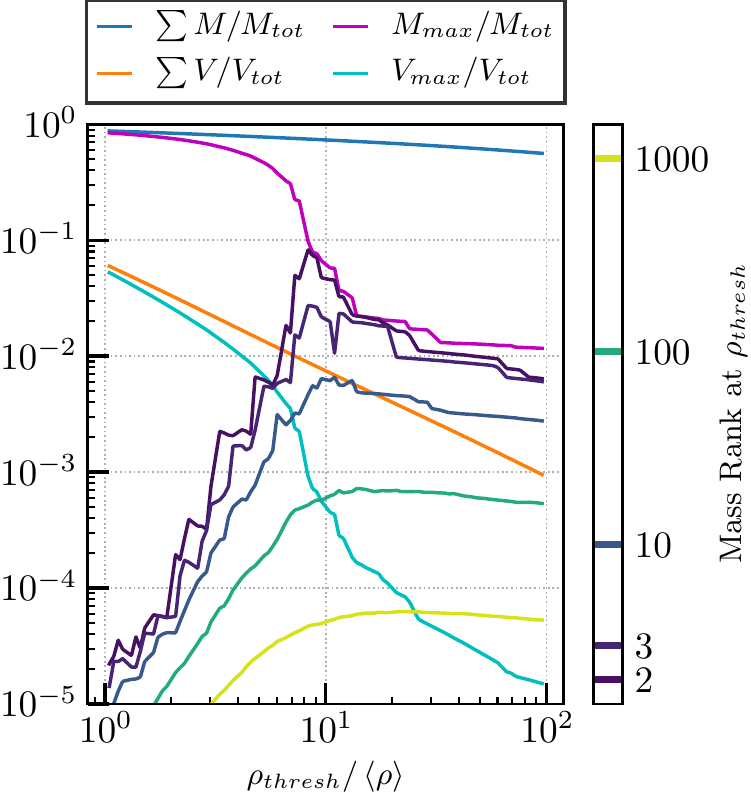}
 \caption{A plot of the same quantities as in \autoref{fig:ms1_perc} but here for the MSII. The smaller box size means that the jump in mass of the largest object is considerably smaller at  percolation than in the MS.}\label{fig:ms2_perc}
\end{figure}

In addition to the summed quantities $\sum M$ and $\sum V$,  \autoref{fig:ms1_perc} and \autoref{fig:ms2_perc} show $M_{max}$ and $V_{max}$, the mass and the volume of the most massive individual object as functions of \rt. For comparison, the masses of the second, third, tenth, hundredth and thousandth most massive objects are also given as functions of \rt. In contrast to the smooth and steady behaviour of the total mass and volume, $M_{max}$ and $V_{max}$ exhibit three distinct regimes, corresponding to two distinct phases separated by a sharp phase transition which occurs as the bounding surface of the largest object percolates. Percolation defines a characteristic density, $\rt = \rper$. Since the percolating object can be considered as the cosmic web, we can say that \rper is the maximum density at which the cosmic web is fully linked.

We first turn our attention to \autoref{fig:ms1_perc}, as the larger MS captures the behaviour in a more pristine way. At densities above \rper, i.e. before the emergence of the infinite cluster, the highly ranked peaks grow exponentially in mass as $\rt$ decreases and they are linked to neighboring objects of similarly high mass. Such exponential growth is typical for systems approaching percolation. Mergers of two or more similarly massive systems are visible as sudden jumps in mass along these curves.\footnote{Note that the identity of the individual objects (as determined by the highest corresponding peak $\pi_m$) may change with \rt along these curves } The accelerating growth in mass is visible even for the 1000\textsuperscript{th} most massive object.

The percolation phase transition happens in the range $6 \lesssim \rt/\rmean \lesssim 7$ and is visible both in $M_{max}$ and $V_{max}$. The mass of the largest connected object increases from below 1\% to over 20\% of the total mass in the simulation for a less than 15\% change in \rt. Its volume increases by the same factor from less than 0.02\% to 0.4\% of the full simulation volume. Immediately after this transition, about one third of the total mass above \rt is contained in the infinite cluster. As \rt decreases further, more and more of the mass and volume at high density are linked to the cosmic web, and $M_{max}$ and $V_{max}$ approach $\sum M$ and $\sum V$, respectively. By the end of the depicted range, at $\rt=\rmean$, about 97\% of the overdense mass is attached to the percolating structure, and very few independent overdensities remain. At this value of \rt, the (logarithmic) difference between $V_{max}$ and the total volume is noticeably larger than the corresponding mass difference, showing that the mean density of the remaining independent objects is considerably lower than that of the web itself. 

As can be seen in \autoref{fig:ms2_perc}, the percolation process in the MSII is less pronounced than in the MS. This is a consequence of its 125 times smaller volume. The mass of the second most massive object maximises at about 8\% of the total simulation mass at $\rt\sim 8\rmean$. As expected, this is similar to the maximum mass of the one hundredth most massive object in the MS and occurs at a similar value of \rt. This leaves little room for the largest object to grow as percolation occurs, and indeed its mass only increases by about a factor of 3, reaching a mass fraction somewhat over 20\%, very similar to the immediate post-percolation mass fraction in the MS. This is much smaller than the jump in mass by a factor of 20 seen at percolation in the MS. The contrast is even starker in the shape of the $V_{max}(\rt)$ curve: in the MSII the volume increases less abruptly, and also less smoothly, than in the MS. This again is a consequence of the limited box size which results in poor sampling of the high-mass tail of the halo mass function and in the largest haloes containing a significant fraction of the total simulation mass. Given these limitations, however, the behaviour agrees well with that in the MS.  Just as in the MS nearly all the mass and volume above $\rt=\rmean$ are part of the percolating object, the cosmic web.

Turning to the $n^{\textrm{th}}$ most massive objects, which are plotted in \autoref{fig:ms1_perc} and \autoref{fig:ms2_perc} for $n = 2, 3, 10,100$ and 1000, we find that all the ranks shown for the MS, and all up to $n=10$ in the MSII, the mass increase accelerates as percolation is approached, just as for the most massive object. Small number statistics introduce some noise in these trends -- the mass of the third-ranked object in the MSII actually decreases just above $\rt=10\rmean$ as some of the highest ranked objects merge. Such noise effects are clearly larger in the MSII because of its smaller volume. This is also responsible for the wider spread in mass between the various curves in the MSII case. While the 1000$^{\textrm{th}}$ most massive object in the MS still corresponds to a rich cluster and is in the exponential tail of the halo mass function, this is not the case in the MSII.

Once percolation has taken place, the masses of all but largest object become smaller and smaller as massive objects are joined to the cosmic web. it. In some cases intermittent increases in mass are visible for the highest mass objects as they connect to other objects in their immediate environment before linking to the web itself, but overall the decline in mass is very rapid. Almost all massive objects are incorporated into the web at \rt values relatively close to the percolation value (see \autoref{sec:abundance} below).


\subsection{The geometry of the Cosmic Web}

\begin{figure*}
 \includegraphics[width=\textwidth]{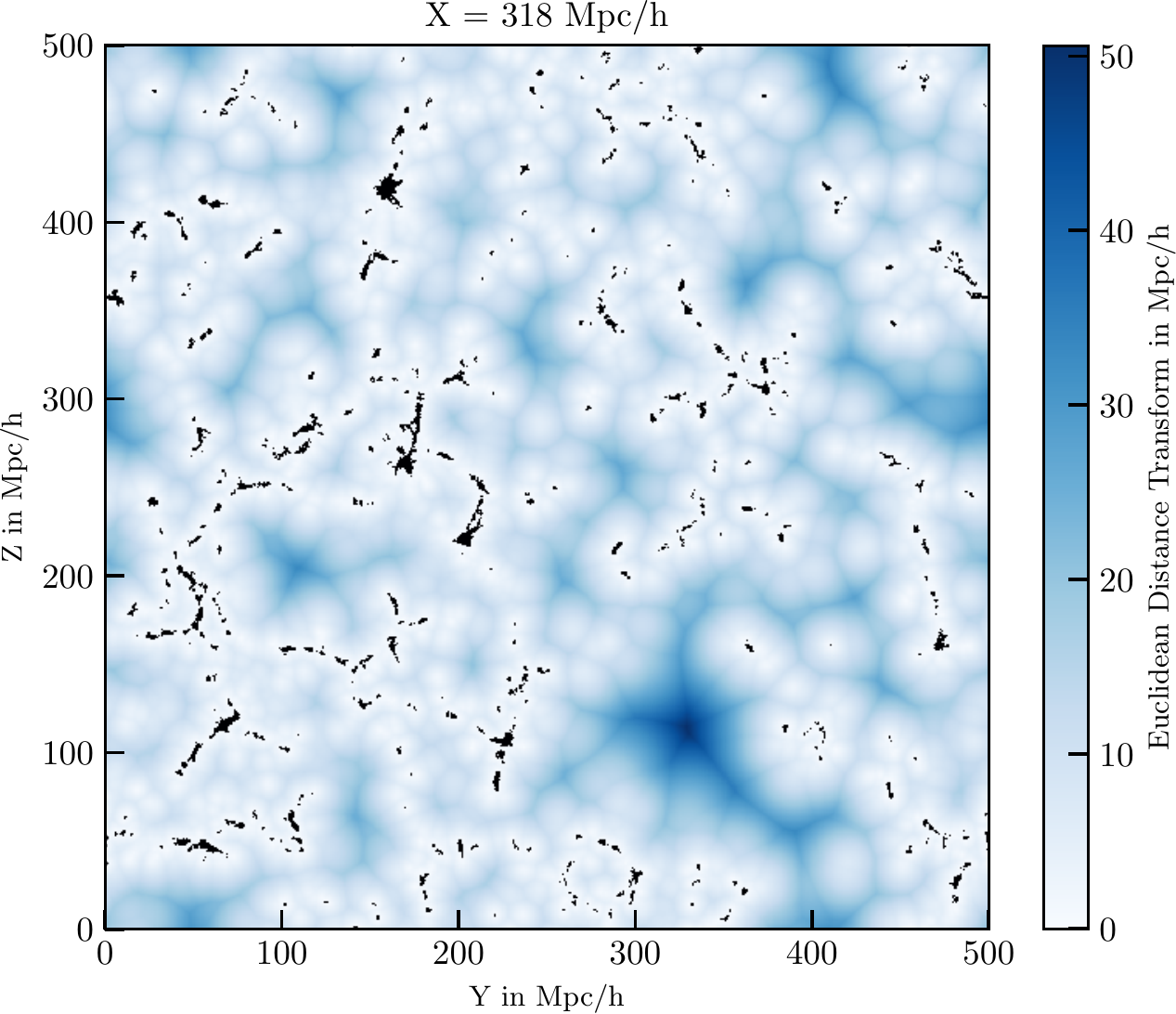}
 \caption{A slice through the MS showing (in black) the percolating object for $\rt = 5.25\rmean$, and (in colour) the Euclidean distance transform (EDT) which gives the minimum 3-D distance from each point to the percolating object. This slice was chosen to contain the global maximum of the EDT, hence the centre of the largest void in the MS for this value of \rt.  }\label{fig:edt_slice}
\end{figure*}

In this subsection we illustrate how to measure characteristic scales for the 
cosmic web, defined as the percolating structure identified by our \methabb. The 
variation of its mass and volume fractions with \rt have been discussed above. 
For the MS case which gives the best statistics, the former increases from 24\% 
to 80\% and the latter from 0.4\% to 7\% as $\rt/\rmean$ drops from 6 to 1 (see 
\autoref{fig:ms1_perc}). In \autoref{fig:edt_slice} we show one layer of a 
$1024^3$ Cartesian grid spanning the full MS volume. We colour black every cell 
that contains at least one particle belonging to the percolating  object for 
$\rt = 5.25\rmean$. (This object contains mass and volume fractions of 35\% 
and 0.62\%, respectively). Clearly, this thin slice intersects the (single) 
percolating object many times. Indeed, defining two black cells in such a slice 
to be part of the same intersection if they share a face, we find that, on 
average, a slice intersects the cosmic web  618 times. Thus the mean distance 
between such intersections is 20.1\Mpch. This characterises the spacing between 
filaments of the web. Given that on average black cells occupy 1.71\% of the 
slice area, the average area of an individual intersection is $(2.63\Mpch)^2$. 
This length scale characterises the thickness of a filament. Since the web 
occupies 1.71\% of the (gridded) MS volume, the total length of web filaments 
within this volume is approximately  $\pi/2~0.0171~(500\Mpch)^3/(2.63\Mpch)^2 = 
\nttt{4.85}{5}\Mpch$, where the factor of $\pi/2$ accounts for the fact that 
filaments intersect a slice like that of \autoref{fig:edt_slice} at random 
angles. These numbers, together with \autoref{fig:edt_slice}, its associated animation, and direct inspection of 3D renderings, confirm that the percolating object is predominantly filamentary. We note, however, that different algorithms produce different identifications of the cosmic web \citep[see][]{libeskind_tracing_2018} so that elements which we classify as part of a filament may be considered as belonging
to sheets or voids by other classifiers, particularly those that employ various characteristic smoothing lengths to define an underlying density field.

An alternative way to characterise length scales is shown by the coloured field 
outside the percolating object in \autoref{fig:edt_slice}. This indicates the 
3-D distance from each cell to the nearest cell which is part of the percolating 
object. This field is called the Euclidean distance transform (EDT) of the 
percolating cell set, and is defined by \begin{equation} E(\v x) = 
\min\limits_{\v w\in \mathcal{W}}\left\|\v x - \v w\right\|, \end{equation} 
where $\mathcal{W}$ is the set of positions of all cells within the percolating 
object and $\|\cdot\|$ denotes the euclidean norm. The local maxima of this 
field give the radii of spheres which are entirely outside the cosmic web but 
touch it at four points. Thus they are locally the largest spherical voids 
within the web. The particular slice shown in \autoref{fig:edt_slice} was chosen 
to contain the highest maximum of $E(\v x)$ (and hence the centre of the largest 
spherical void) in the MS volume, for which 
$R_\mathrm{void,max}\approx\SI{50.5}{\Mpch}$.\footnote{An animation of this plot 
illustrating the 3D structure of the percolating object and of the EDT can be 
found at \url{https://pbusch.net/tlt/perc_movie.mp4}} 

\begin{figure}
 \includegraphics{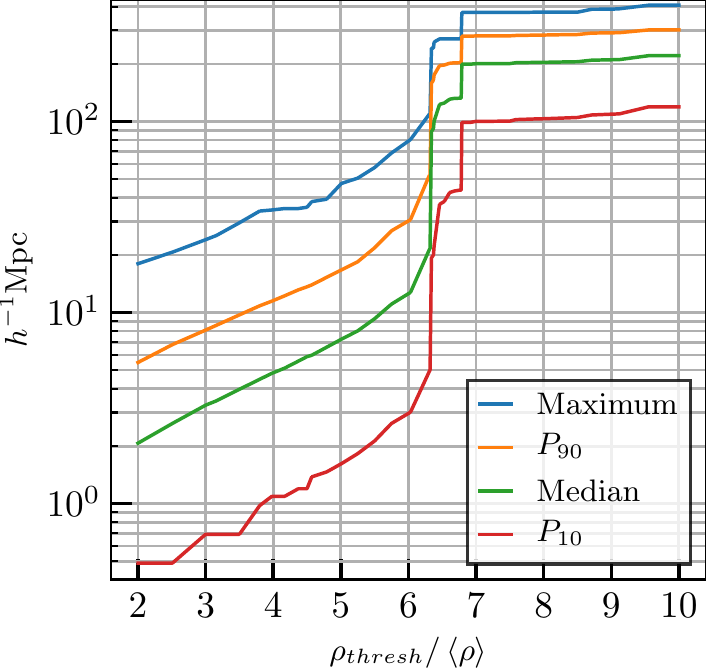}
 \caption{Variation of the distance distribution derived from the Euclidean Distance Transform (EDT) of the space external to the largest connected object in the MS as a function of the threshold density \rt of its bounding surface. The blue curve shows the maximum value of the EDT (i.e. the radius of the largest spherical void) while orange, green, and red curves give, respectively, the 90\%, the median and the 10\% points of the distance distribution at each value of \rt. Percolation is evident in the abrupt jump in these curves at $\rt/\rmean=6.3$. The smaller jump at somewhat higher threshold is due to the merging of two objects of nearly similar size and corresponds to the lowest density at which the first and second most massive objects in \autoref{fig:ms1_perc} are of similar mass.}\label{fig:edt_percs}
\end{figure}

The properties of $E(\v x)$ can be used in many ways to to quantify the cosmic 
web. Here, we restrict ourselves to a simple example and to a test of 
convergence between our two simulations. In \autoref{fig:edt_percs} we show how 
the distribution of $E(\v x)$ varies with $\rt/\rmean$ in the MS, plotting the 
10\%, 50\% and 90\% points of the distribution, as well as its maximum. At the 
largest threshold shown,  $R_\mathrm{void,max}$ is already significantly below 
its maximum possible value $\sqrt{3}\times 250\Mpch = 433\Mpch$, showing that 
the most massive object is much larger than an individual halo. As \rt is 
lowered $R_\mathrm{void,max}$ initially decreases slowly, but then drops 
precipitously to about $100\Mpch$ as percolation occurs in two steps over the 
narrow range $6.3 < \rt/\rmean<6.8$. As \rt is reduced further 
$R_\mathrm{void,max}$ continues to decrease steeply, reaching a value of about 
20\Mpch for  $\rt/\rmean=2$, the smallest threshold plotted.

The median and the upper and lower decile points of the distance distribution vary with \rt in a qualitatively similar way to its maximum value, but there are some notable systematic differences. The jump across the percolation transition varies substantially, from a factor of 3.5 for $R_\mathrm{void,max}$ to factors of 6.5, 9.8 and 12.7 for the 90\%, 50\% and 10\% points, respectively. This reflects a broadening of the distance distribution which continues more slowly as \rt decreases further. The 10\% and 90\% points differ by factors of 2.8, 10.2, 10.3 and 11.8 for  $\rt/\rmean = 7, 6, 5$ and 3, respectively. This change in shape is a consequence of the change in geometry from a single relatively compact object for $\rt/\rmean>9$ to a volume-filling network of filaments for $\rt/\rmean<6$. The substantial, quasi-exponential drop (from 12 to 2.3\Mpch) in the median distance as $\rt/\rmean$ decreases from 6 to 2 is due to the growth of low-density filaments which extend from the higher density web  into previously empty regions. Over this range, the total length of filaments in the MS (estimated as above) increases from \nttt{2.63}{5}\Mpch to \nttt{2.65}{6}\Mpch. Even for $\rt/\rmean=2$ the percolating object fills only about 5\% of the total volume, but the remaining 95\% of the simulation is much more densely threaded with filaments than in \autoref{fig:edt_slice}.

\begin{figure}
  \includegraphics{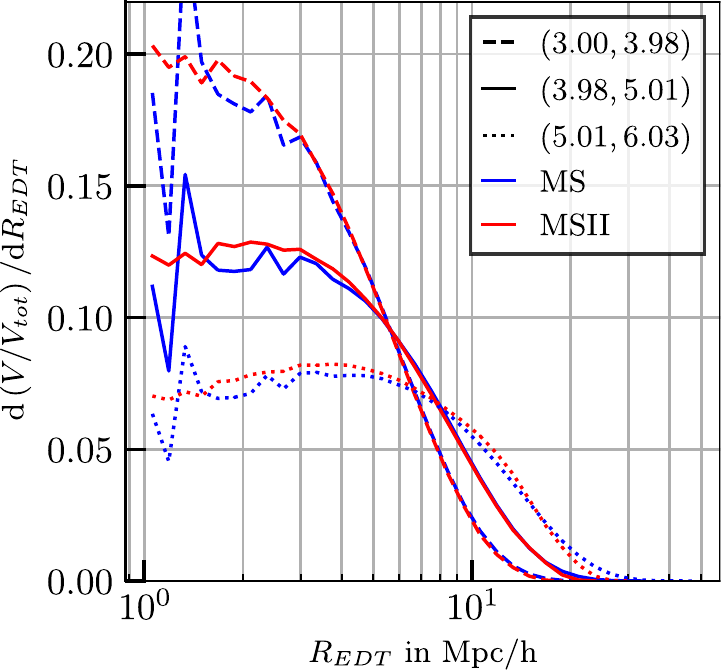}
  \caption{The EDT distance distribution relative to the percolating object at threshold densities of $\rt\approx\{3,4,5\}\rmean$ and $\rt\approx\{4,5,6\}\rmean$ in the MS and the MSII, respectively. At these thresholds the cosmic web contains mass fractions of $\{60,50,40\}\%$ in both simulations.}\label{fig:dist_vol_comp}
\end{figure}

As seen in \autoref{fig:ms1_perc} and \autoref{fig:ms2_perc} percolation occurs at a higher density threshold in the MSII than in the MS, and the curves showing the mass and volume of the percolating object are both shifted slightly to the right in the MSII case. As a result it is not clear how best to check for convergence of web properties between the two simulations.  In \autoref{fig:dist_vol_comp} we compare the EDT distance distributions obtained when the thresholds in the two simulations are matched in such a way that they produce the same mass fraction in the percolating object. Specifically, we choose thresholds $\rt/\rmean \approx \{3,4,5\}$ and $\{4,5,6\}$ in the MS and MSII respectively, which leads to mass fractions of $\{60\%, 50\%, 40\%\}$ in the percolating object in both simulations. With this choice, the distance distributions agree remarkably well (apart from some small-scale discreteness effects in the MS) despite the difference in mass resolution of a factor of 125 and the change by a factor of two in the median of the EDT distribution over this range of thresholds. The longer tail to large distances in the MS for $\rt/\rmean = 5$ is clearly a reflection of its much larger volume; voids of $100\Mpch$ diameter would not fit in the MSII simulation box. Most properties of the cosmic web as characterised by the EDT distance distribution are clearly very well converged in the Millennium Simulations.

\section{The Abundance of Peaks}\label{sec:abundance}

  The abundance of peaks as a function of their mass, usually known as the mass function, is perhaps the simplest characterisation of the population of objects defined by our \method. We begin by using our two simulations to compare the masses of thresholded objects, limited at various thresholds, to those of the corresponding haloes defined by the standard FoF algorithm. For the threshold which leads to the closest correspondence, we then differentiate the set of thresholded objects by their \rl values to understand the local environments in which they live. We close by conducting similar investigations for unthresholded objects. Throughout this section we express abundances as number densities in units of $\si{\per\h\cubed\per\Mpc\cubed}$. For the MS a single object implies an abundance of $n=\nttt{8}{-9}\si{\per\h\cubed\per\Mpc\cubed}$. For the MSII the minimum number density is $n=\ttt{-6}\si{\per\h\cubed\per\Mpc\cubed}$.

  \subsection{Thresholded Peaks}\label{sec:abund_th}
  We begin by discussing thresholded peaks (persistence filtered and including the mass of their subpeaks) because, as we will show, these correspond quite closely to the usual definition of haloes in cosmological simulations.
    
    \subsubsection{The FoF-TLT Correspondence}\label{sec:massf}

      We consider a thresholded peak and a FoF halo (identified using $b=0.2$) to be one and the same object if the peak particle of the group is a member of the FoF halo \emph{and} the most bound particle of the FoF group is part of the thresholded peak. Since the FoF catalogues for our simulations are limited to halos with at least 20 particles, we only consider thresholded peaks with at least 100 particles in order to be sure that their counterparts cannot fall below the catalogue limit. Motivated by the analysis in \cite{more_overdensity_2011}, we consider thresholds from the set $\rt/\rmean\in\{60,80,100,125\}$ (more precisely \ttt{\{1.8,1.9,2,2.1\}}). These authors predict a number and resolution dependence of the bounding density of FoF objects which nicely reproduces
      their numerical data, and this range should cover the values they find to correspond to FoF linking parameter $b=0.2$. For these four thresholds, 99.91, 99.85, 99.55 and 98.58\% of the thresholded peaks with more than 100 particles are matched to a FoF halo by the above criteria. Most of the failures are due to the FoF algorithm linking the particles of the thresholded peak to a higher peak which is still disjoint according to the TLT algorithm.
      
\begin{figure}
	\centering
	\includegraphics[scale=1.]{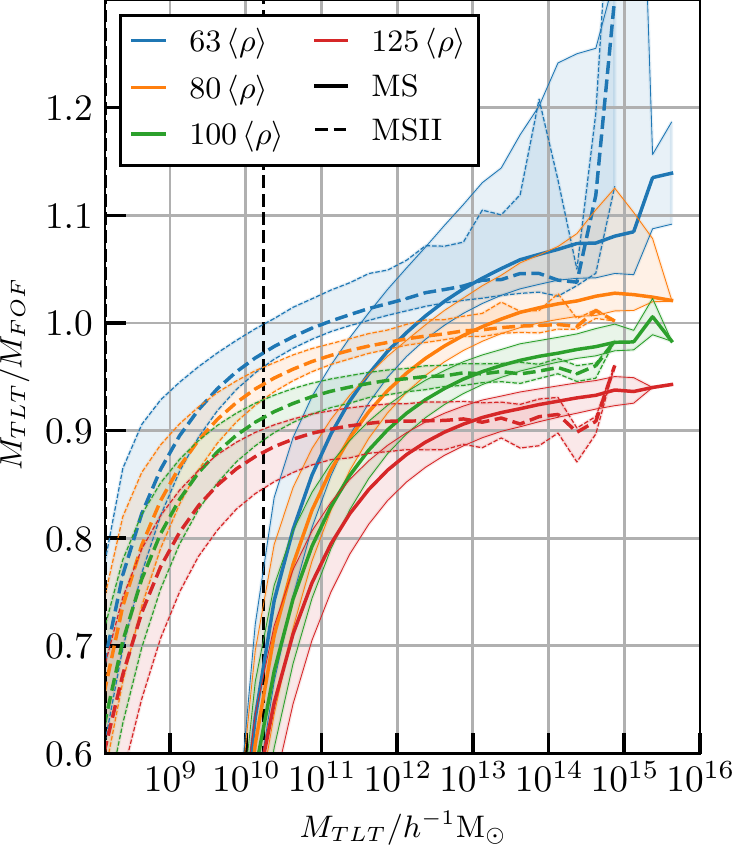}
	\caption{The median and the 16 to 84\% range of the ratio of the mass of a TLT peak to that of its corresponding FoF halo as a function of TLT peak mass. Solid curves refer to the MS, dashed curves to the MSII, with different colours giving results for four different thresholds as given in the legend. The turn-down in each set of curves at low mass is a surface-to-volume effect which becomes significant as the number of particles drops below about 1000. For $\rt\approx 80\rmean$, well resolved TLT peaks are assigned about the same mass as the corresponding FoF halo with rather small scatter.}\label{fig:mf_rats}
      \end{figure}

      For the sets of matched objects, \autoref{fig:mf_rats}  shows the median and scatter in the ratio of the thresholded peak mass to the FoF halo mass as a function of the former.
      The medians of these ratios vary more strongly with threshold at higher mass as a consequence of the lower concentration of high-mass haloes. For each threshold, the median value of the ratio decreases with decreasing mass, consistent with the trends found by \cite{more_overdensity_2011}.  This systematic difference between FoF and TLT arises primarily from the treatment of the outermost particles. For a particle to be part of a FoF-halo it is sufficient for it  to be closer than the linking length to another halo particle. To be above threshold for inclusion in a TLT peak, however, the distribution of particles outside the peak is also important, since they determine the extent of the particle's Voronoi cell. This difference results in an offset between the masses found by the two algorithms which depends on particle number through a surface-to-volume effect; details of the treatment of the outermost particle layer decrease in importance as the total number of particles increases.
      
      For masses corresponding to at least about 1000 particles in the MS, there is good numerical convergence between the median TLT-FoF halo mass ratios found in our two simulations. At lower masses the stronger surface-to-volume effect in the MS causes the results to diverge. The relatively small scatter in the mass ratio at all masses, together with the small fraction of TLT peaks that do not have a unique FoF halo counterpart according to our criteria, shows that the two algorithms identify very similar sets of haloes.\footnote{We also find that a similarly small fraction of FoF haloes do not correspond to a unique TLT peak.} For $\rt=80\rm$ and large enough particle number, TLT peaks are typically assigned the same mass as the corresponding FoF haloes (for $b=0.2$)  with a scatter of only about 5\%. 
      
    \subsubsection{Distribution in \texorpdfstring{$M$-$\rho_{lim}$}{F_sub in M-rho\_lim}-Space}
    
\begin{figure}
        \centering
        \includegraphics[width=\linewidth]{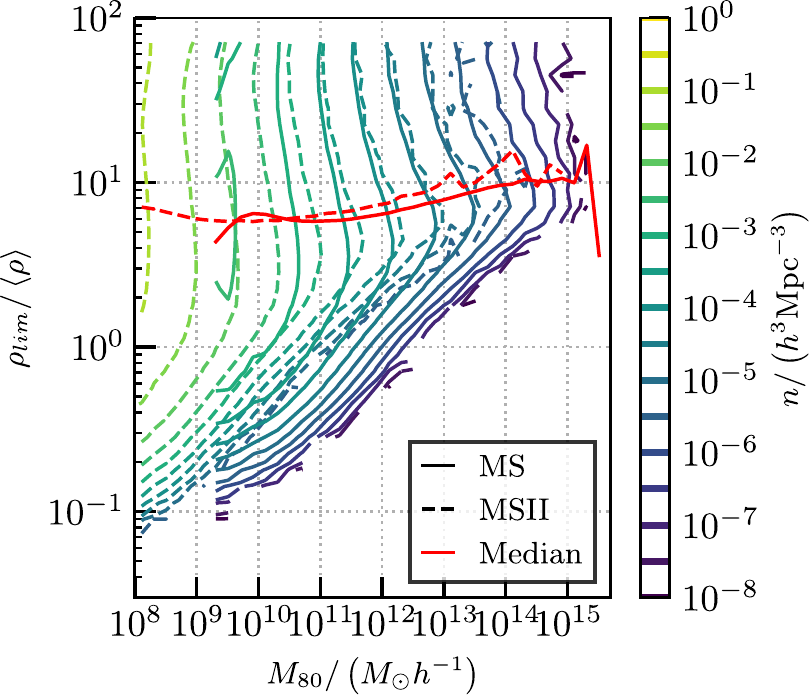}
        \caption{The abundance of thresholded peaks in the MS (solid) and MSII (dashed) as a function of their mass and their \rl value. Coloured lines show logarithmically spaced equi-abundance contours as indicated by the sidebar. Red lines indicate the median values of \rl at each mass. There is good agreement between the simulations for well resolved ($N_{part}>100$, $\sim\ttt{11}\si{\Msolh}$ for the MS) objects with \rl above the median value. Agreement is less good at lower \rl values. }\label{fig:comb_thresh_abundance}
      \end{figure}

      The \methabb data structure provides an additional property for thresholded peaks that is not available for the objects defined by other halo-finders, namely the limiting density \rl. Since this is a measure of the immediate environment of the halo, it is interesting to see how haloes of a given mass are distributed in \rl. In \autoref{fig:comb_thresh_abundance} we show the joint distribution over mass and \rl of TLT peaks thresholded at $\rt=80\rmean$.  The number densities of peaks in the MS and the MSII are compared by superimposing their sets of logarithmically spaced isodensity contours in $\log M$-$\log \rho_{lim}$-space. We also show (as red lines) the median value of \rl at each mass.
      
      The first interesting result from \autoref{fig:comb_thresh_abundance} is that the median value of \rl is almost independent of peak mass, is well converged between the two simulations, and is close to but somewhat above the density at which percolation occurs (see \autoref{sec:perc}). Thus almost half of all haloes are not part of the cosmic web as we defined it above, with this fraction declining slowly with increasing mass.
   
      The isodensity contours for the two simulations are qualitatively similar, but with some noticeable differences in detailed structure. Above the median value of \rl there is good agreement when both have good statistics and resolve the haloes adequately (between about $\ttt{11}\si{\Msolh}$ and $\ttt{13}\si{\Msolh}$). In this regime the contours are nearly vertical, indicating that, at fixed mass, abundance varies only slowly with  \rl. Abundances
      peak about a factor of 2 below the median value of \rl and then decline rapidly at lower values. This decline reflects the rapid increase in the mass of the percolating object seen in \autoref{fig:ms1_perc} and \autoref{fig:ms2_perc} as the threshold drops below the percolation value. The main difference between the two simulations is that this decline is significantly faster in the MSII than in the MS, as can be inferred also from the two earlier figures where the percolating object grows to large fractional mass noticeably earlier in the MSII case. This reflects, perhaps, that the low-mass filaments which connect these ``isolated" haloes to the cosmic web are better resolved in the MSII.
    
  \subsection{Unthresholded Peaks}\label{sec:abund_unth}
 
    Unlike the thresholded peaks of the previous section, unthresholded peaks do not have a direct correspondence in the usual halo picture. We therefore look only at their abundance in $\log M$-$\log \rho_{lim}$ space and use this to investigate the effect of persistence filtering by comparing the distributions for filtered and unfiltered peaks. In contrast to the last section, we here do not include the mass of a subpeak in that of its parent. Thus the sets of peaks considered are strict partitions of all simulation particles into disjoint subsets.

\begin{figure}
        \includegraphics[width=\linewidth]{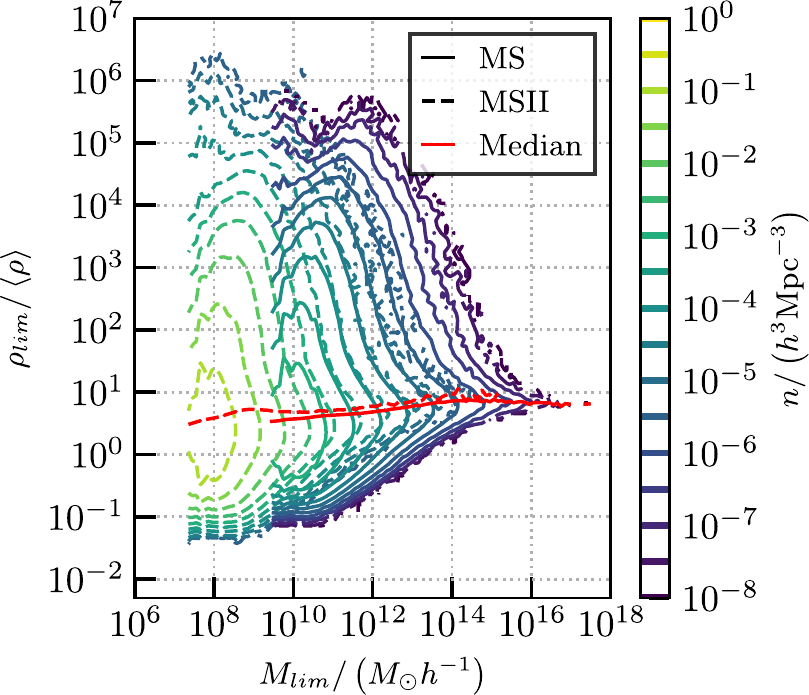}
        \caption{The distribution of persistence-filtered unthresholded objects over mass and limiting density. Due to volume constraints the MSII is restricted to abundances above $\ttt{-6}\si{(\Mpch)}^{-3}$. Red lines indicate the median value of \rl at fixed mass. For these medians the two simulations agree very well at all masses and there is at most a weak trend with mass. The contours also agree well for masses larger than $\ttt{11}\si{\Msolh}$. The drop in abundance at lower masses in the MS is due to the removal of objects by persistence filtering. A similar change in contour shape is visible for the MSII at 125 times lower mass.}\label{fig:comb_unthresh_abundance}
        \end{figure}

\begin{figure}
        \includegraphics[width=\linewidth]{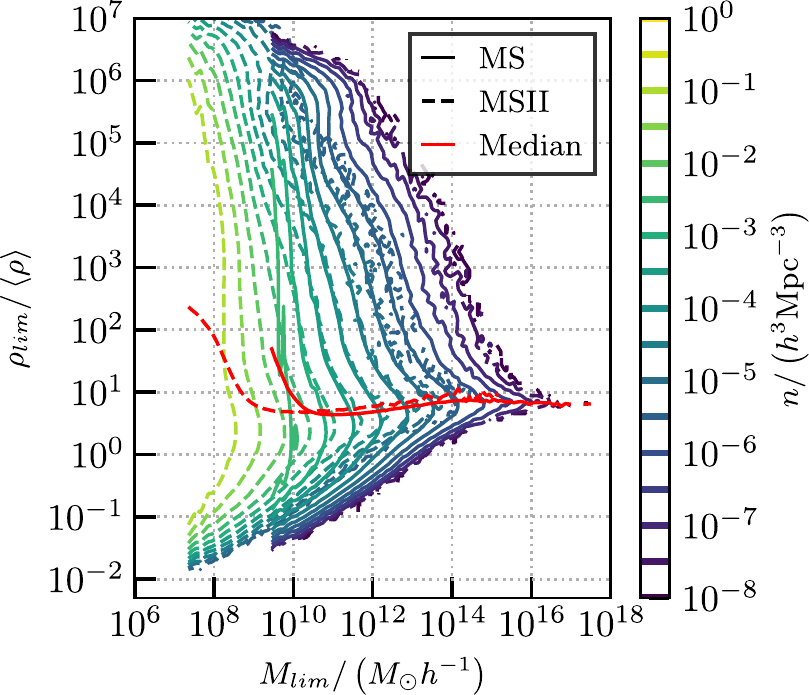}
        \caption{As \autoref{fig:comb_unthresh_abundance} but without any persistence filtering.
        The two simulations again agree very well at masses above a few times $\ttt{10}\si{\Msolh}$. At lower masses there is now, however, an excess of objects in the MS, due to the presence of a significant number of ``noise" peaks. These noise peaks occur primarily at large \rl, causing an substantial turn-up in the median curves at the relevant masses.}\label{fig:comb_all_unthresh_abundance}
        \end{figure}

    To compare abundances in the MS and the MSII and to isolate the effects of persistence filtering, we show contours of constant abundance in  \autoref{fig:comb_unthresh_abundance} and \autoref{fig:comb_all_unthresh_abundance}. The format here is identical to \autoref{fig:comb_thresh_abundance} except now with many more orders of magnitude in \rl on the vertical axis. These two plots differ only in the application of a persistence filter (with a density ratio of 10) for the first but not for the second. They are essentially identical for masses above $\sim\ttt{11}\si{\Msolh}$ and \rl values below $10^4\rmean$, showing that persistence filtering has little effect in this regime. 
        
    The median curves for \rl as a function of peak mass agree well between the two simulations and depend only weakly on mass in the case with persistence filtering. Without filtering, however, these curves turn up sharply at a peak mass corresponding to several tens of particles in each simulation. This shows that spurious peaks due to discreteness noise become significant at these masses and are present primarily at large \rl where they would potentially be considered as subhaloes of more massive objects.
        
    Wherever statistics for the MSII are  sufficient to judge, the agreement between the two simulations is good for peak masses above $\ttt{11}\si{\Msolh}$. Thus the statistical properties of our TLT structure appear robust and unaffected by discreteness noise in this regime. At lower masses the effect of this noise and of our persistence filter can be assessed by comparing contours for the MS and MSII. When the filter is applied (\autoref{fig:comb_unthresh_abundance}), the contours for the MS change sharply in slope and start to deviate from those for the MSII at a peak mass which increases from $\sim 2\times\ttt{10}\si{\Msolh}$ at $\rl \approx 30\rmean$ to $\sim \ttt{11}\si{\Msolh}$ at $\rl\approx 10^4\rmean$. This is the result of the exclusion of real structures of lower mass by the persistence filter. In contrast, without persistence filtering (\autoref{fig:comb_all_unthresh_abundance}) the contours for the MS rise above those for the MS at peak masses below several times $\ttt{10}\si{\Msolh}$, reflecting the presence of the spurious structures which also pull up the median of \rl. 
        
    At low values of \rl the contour shapes differ between the two simulations in the same way as in
    \autoref{fig:comb_thresh_abundance}, and presumably for the same reason -- weak sheets and filaments are better resolved in the MSII than in the MS. Persistence filtering also introduces a noticeable change in shape in this regime, reducing the slope of the contours at masses corresponding to 100 particles or fewer. Clearly many of the low-mass structures identified with low values of \rl have a weak contrast (corresponding, perhaps to segments of filaments rather than to haloes) and so are eliminated by the persistence filter. 
        
    A final noticeable difference between the distribution for thresholded peaks (\autoref{fig:comb_thresh_abundance}) and those for unthresholded peaks (\autoref{fig:comb_unthresh_abundance} and \autoref{fig:comb_all_unthresh_abundance}) is the greater prominence of the abundance ``spike" close to the median value of \rl. This is easily understood as a result of the joining of individual haloes into ``superclusters" as \rl approaches the percolation value.

       From the results in this section we conclude that a simple persistence filter does indeed help with removing spurious peaks but it also removes a significant number of real low-contrast features. For the purposes of this work it performs reasonably well, and as long as we ensure convergence with the well resolved region in the MSII or we deal with the high-mass regime of the MS we can trust our results. In practice, this means that results based on structures with more than 100 or so particles appear to be robust
      
\section{Density-Mass Profiles for Peaks}\label{sec:mass_dens_profs}

As a first application of the Tessellation Level Tree to characterise the internal structure of density peaks, we here briefly discuss the density-mass profiles we motivated and described in \autoref{sec:den_prof}. 

\begin{figure*}
    \includegraphics[scale=1.]{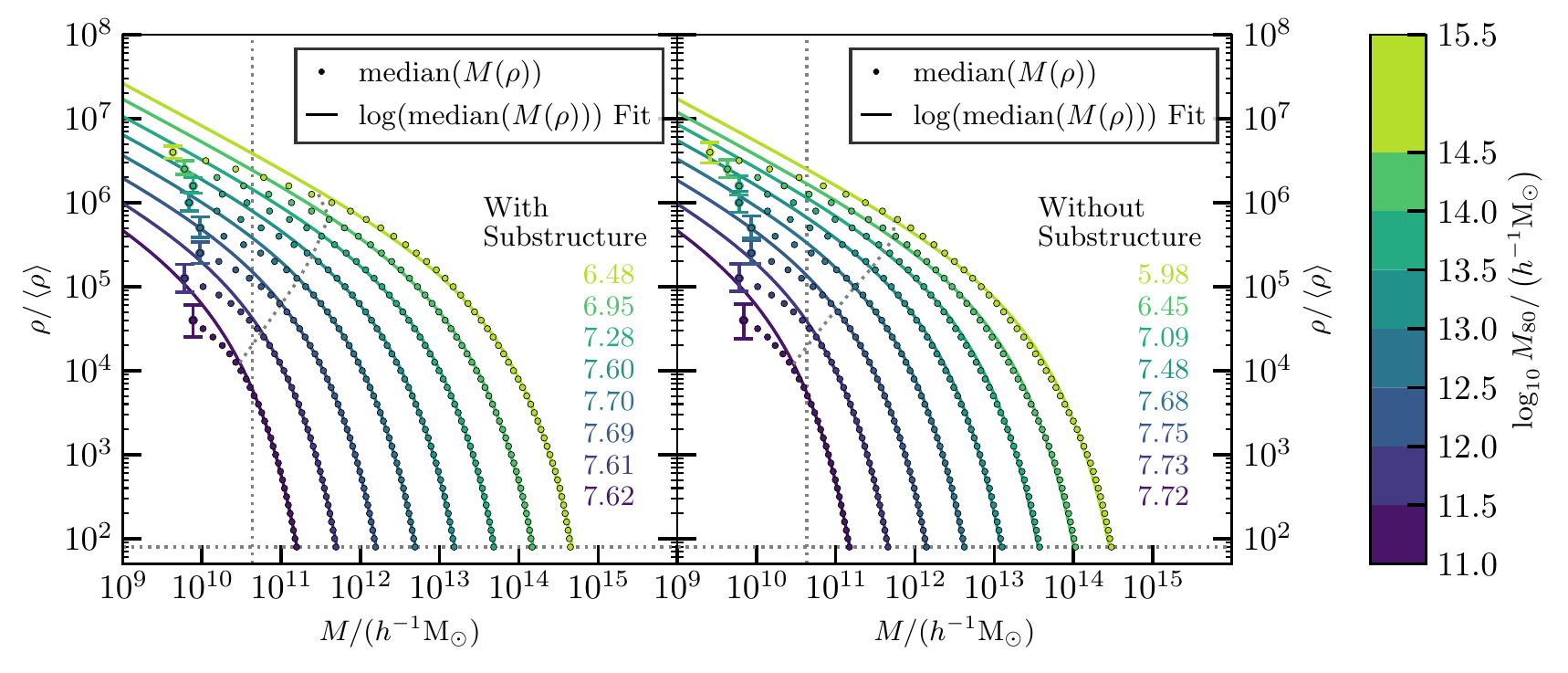}
    \caption{Median density of the equidensity surface with given enclosed mass as a function of that mass for thresholded peaks binned in $M_{80}$, as indicated by the colour bar at the right. The left panel includes the mass in subpeaks (TH\scp(80)) while the right panel only includes mass directly under the main peak (TH\scm(80)). The bar on the lowest enclosed mass plotted for each profile shows the interquartile range in the equidensity values for the peaks in that $M_{80}$ bin.  The vertical grey dotted line corresponds to an enclosed mass of 50 particles, while the diagonal grey dotted line corresponds to 3 gravitational softening radii. The smooth curves are NFW profiles fit to the points to the right of these lines and above the horizontal grey dotted line marking the threshold defining the peaks. Coloured numbers in each panel indicate the values of the NFW concentration $c_{200c}$ for the corresponding fits.}\label{fig:mass_dens_profs}
  \end{figure*}

In \autoref{fig:mass_dens_profs} we present profiles as a function of peak mass $M_{80}$ for thresholded peaks in the MS, both including (TH$^+$(80), left panel) and excluding (TH$^-$(80), right panel) the mass of subpeaks. Circles indicate the median value of $\rho(M)$, the density on the equidensity contour enclosing mass $M$, for simulation peaks with thresholded mass $M_{80}$ lying in the ranges indicated by the colour bar to the right of the plot. Solid lines show fits using a reformulation of the NFW profile \citep{navarro_universal_1997} as a function of enclosed mass instead of radius. The bars on the lowest mass plotted for each $M_{80}$ bin indicate the interquartile range of the densities at that enclosed mass for peaks with $M_{80}$ in the relevant bin. Only points to the right of the two grey dotted lines (corresponding to 50 particles (vertical) and to three times the MS gravitational softening radius (slanted)) are used in the fit. The lower bound in density is set at the density threshold defining the peaks. We fit the profiles to obtain the original NFW-parameters $\delta$ and $r_s$ using the expressions for $\rho(M)$ derived in the Appendix. The corresponding values of NFW concentration ($c_{200c} = R_{200c}/r_s$) are indicated in the appropriate colour in each panel.

We find that both cases are very well fit by NFW-profiles over the fitting range we consider. At smaller enclosed mass (and hence radius), we see a systematic departure towards lower densities as gravitational softening limits the attainable central densities. For high-mass haloes, the NFW concentration values are slightly larger in the case
with substructure, while in the opposite is true for low-mass haloes. This reflect the fact that subhaloes tend to have larger
typical densities than the main halo in well resolved objects, but smaller ones when particle discreteness is a limiting factor. In both cases, for well resolved haloes the concentrations are larger than found by fitting the spherically averaged radial density profiles of MS haloes \citep{neto_statistics_2007}. This is because the averaging smooths out the subpeaks and adds their mass preferentially at large radii, hence at relatively low (averaged) density.

Interestingly, for massive objects the profile including substructure is smoother and better fit by the NFW formula than the one excluding it, which has a slight deficit in mass at intermediate densities and an excess at high densities. This reflects the fact that the  highest densities resolvable in a halo are typically at the centre of its main component, whereas subhaloes contribute significantly at densities which are higher than the bounding value but typically well below the peak value. It is notable (and convenient) that the profiles including substructure are so well fit by the simple two-parameter NFW formula, since these profiles make no assumption about the spatial or geometrical arrangement of the mass at each value of the local density $\rho$,and so can be used to make analytic estimates of the total dark matter annihilation luminosity of haloes, including the effects of substructure and other deviations from spherical symmetry.

\section{Clustering of Peaks: Assembly Bias}\label{sec:ab_res}
  
   The limiting density \rl measures the density at which a peak is linked to a more massive structure, and is hence an indicator of its immediate environment. As a first example of how the TLT can be used to study the statistics of large-scale structure, we now  use the methods of \autoref{sec:bias_def} for a brief discussion of how the clustering of density peaks is biased as a function of \rl.
   
\begin{figure}
    \centering
        \includegraphics[scale=1.]{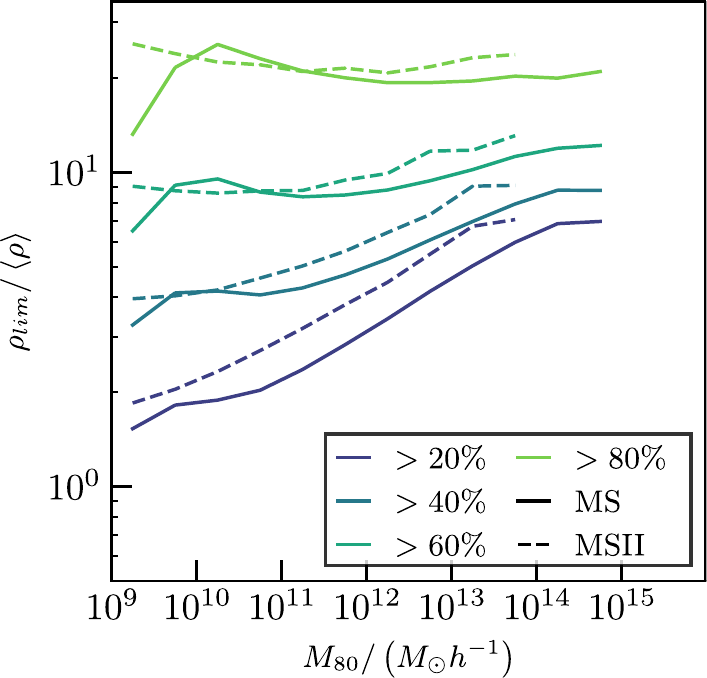}
    \caption{Boundary values between the quintiles in \rl in 0.5 dex bins in $M_{80}$ for the MS (solid) and the MSII (dashed).}\label{fig:perc_bounds}
  \end{figure}

\begin{figure}
    \centering
        \includegraphics[scale=1.]{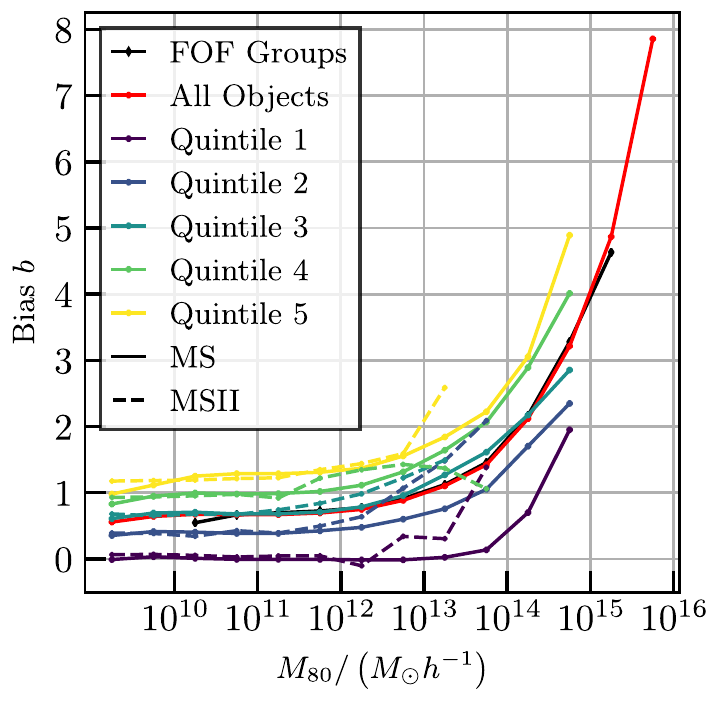}
    \caption{Bias for the quintiles in \rl in mass-binned subsamples of thresholded haloes with $\rt = 80 \rmean$ in the MS and MSII. The red line gives the bias of the mass-binned subsamples not split by \rl in the MS and the black line that of the central subhalo in FoF groups binned by FoF group mass. Only bins with at least 100 members are shown. }\label{fig:bias_mass_perc}
  \end{figure}

   In order to compare with previous work which analysed assembly bias in the MS using FoF haloes \citep{gao_assembly_2007,faltenbacher_assembly_2010} we consider thresholded peaks (including subpeaks, i.e. TH$^+$(80)) and divide them into bins of width \SI{0.5}{\dex} in $M_{80}$. The peaks in each bin are then split into five equal subsamples according to their \rl values. The values of \rl separating these quintiles are shown as a function of $M_{80}$ in \autoref{fig:perc_bounds}. Consistent with the results found in \autoref{fig:comb_thresh_abundance}, the values for the MSII are higher than for the MS for most $M_{80}$ values.  We then calculate the large-scale clustering bias $b$ for each subsample as described in \autoref{sec:bias_def}. The results in \autoref{fig:bias_mass_perc} show a very strong dependence of $b$ on \rl, indicating stronger assembly bias than for any known internal property of haloes. For the MS, we also present bias values for all the objects in each mass bin; these agree perfectly with those found previously for FoF haloes. Results for the two simulations agree well at low mass, but become noisy for the MSII above about $\ttt{12}\si{\Msolh}$. This is despite the fact that we only only show results for bins with at least 100 members in total in an attempt to reduce uncertainties in the halo-matter cross correlations.
    
  The differences in bias are so large that the lowest quintile is \emph{uncorrelated} with the large-scale matter distribution for masses $M_{80}\leq\ttt{13.3}\si{\Msolh}$. Choosing a lower cut on \rl, for example, taking the bottom decile, actually leads to a negative bias, i.e. to anticorrelation with the large-scale density field.  Bias values are nearly constant below about $\ttt{12}\si{\Msolh}$, but start to rise substantially at higher masses. This rise begins at lower masses for the higher quintiles and becomes appreciable for the lowest quintile only above $\ttt{13.5}\si{\Msolh}$. As a result, the spread in bias at given peak mass increases with peak mass. 
  
  This behaviour reflects a mass-dependent relation between peaks and the cosmic web. Lower mass objects, corresponding to galaxy and group haloes, have values of \rl that extend well below that needed for percolation (see \autoref{fig:comb_thresh_abundance}); such low values correspond to objects that are not part of the percolating structure. At high masses, however, this is not the case, and almost all cluster-mass peaks have \rl values  sufficient to attach them to the cosmic web as defined in \autoref{sec:above}.  This is seen clearly in \autoref{fig:perc_bounds}; while the two upper inter-quintile boundaries vary  little with peak mass, the lower ones, particularly the 20\% point of the distribution, rise quite strongly towards high mass. 
  
  In a forthcoming paper, we will investigate in more detail the strong assembly bias we find as a function of \rl, analysing its relation both to the structure of the cosmic web, and to assembly bias as a function of various internal properties of peaks.

\section{Summary and Conclusions}


We present a new data structure which we call a ``\method'' (\methabb). This is defined on a cosmological N-body simulation using the densities and neighbour relations associated with each particle by a Voronoi tessellation of the full particle distribution. The \methabb  partitions the particles into disjoint subsets, each associated with a single local density peak. These peaks are then associated in a tree structure which links each one with a unique higher peak of which it can be considered a subpeak. The minimum density associated with a peak is slightly above the density of the ``saddle-point" particle linking it to its higher density parent. We call the latter the peak's limiting density \rl. It provides an unsmoothed measure of the peak's environment, which supplements other more conventional properties such as mass, volume, shape, spin, density profile, etc. Spurious peaks due to discreteness noise can be removed by a persistence filter requiring the ratio of peak to limiting density, $r\geq10$, although such filtering also removes many real low-contrast peaks.

We additionally introduce the concept of thresholded peaks which are defined as connected sets of particles, all of which have densities exceeding some chosen threshold \rt. This definition is quite close to that of standard ``friends-of-friends" (FoF) haloes. A 2D topographical analogue would be islands on a map, where the threshold corresponds to sea level. Such thresholded peaks have properties analogous to those of FoF haloes (mass, shape, spin...) but the \rl value of their highest peak (which by definition must be less than \rt) provides a novel measure of the environment of the peak/halo. 

As a first application of the \methabb we looked at the total mass of all thresholded peaks and at the mass of the single most massive one as a function of threshold, \rt.  While in both the MS and the MSII the total mass increases smoothly with decreasing threshold, the mass of the largest structure grows very rapidly at a percolation transition which occurs at $\rper \approx 7\rmean$ in the MS and at $\rper \approx 9\rmean$ in the MSII. At lower thresholds the largest object can be considered as defining the cosmic web and the mass of the second and lower ranked objects decreases rapidly as they progressively join it.  The MSII percolates at slightly higher density because its smaller particle mass allows weaker high-density filaments to be resolved. Percolation is also less pronounced in the MSII, as its smaller volume results in a lower contrast between the mass of the web and that of the largest isolated peaks. In both simulations, percolation sets in when thresholded peaks account for a volume fraction $V(\rho\geq\rper)/V_{tot}\approx 0.01$ and a mass fraction  $M(\rho\geq\rper)/M_{tot}\gtrsim 0.7$, hence have a mean overdensity of about 70.

A different picture of the percolation process is provided by the Euclidean Distance Transform (EDT), the distribution of distances from random points outside the largest connected structure to the closest point within the structure. This distribution changes shape at percolation as the largest object shifts from being of finite size to being space-filling. For $\rt<\rper$ the EDT characterizes the size distribution of the low-density regions enclosed by the web, and it is remarkable that when parametrised by the volume fraction occupied by the web, it is very well converged between the MS and the MSII despite the factor of 125 in mass resolution between the two simulations.


We compared thresholded peaks in our two simulations with FoF haloes defined using the standard linking length $b=0.2$. For appropriately chosen \rt ($\approx 80$) the correspondence is very good and the great majority ($>99\%$) of \methabb peaks with mass of more than 100 particles can be unambiguously associated with a FoF halo with very similar mass; the scatter in the mass ratio is only about $5\%$. 


We also introduce ``density-mass profiles" as a new tool for studying the density structure of peaks and haloes in N-body simulations. These profiles plot density against mass, where at each density the mass is the total for all particles with (Voronoi-estimated) density exceeding that value. As mass increases from zero to that of the halo, density drops from its peak value to \rt. This construction has some advantage over traditional spherically averaged  radial density profiles. There is no smoothing, no geometrical assumptions are made, and the profile can be used straightforwardly to estimate the annihilation luminosity of a halo accounting for all substructure resolved by the simulation. These profiles turn out to be well fit by simple NFW formulae, although with slightly different parameters as a function of halo mass than found for standard radial density profiles. 


Finally, we investigate the large-scale clustering bias of thresholded peaks as a function of limiting density \rl. At given peak mass, clustering varies very strongly with \rl. Indeed, for FoF-like objects, the assembly bias effect we find for this quantity is stronger than that for any known internal property of haloes, and is comparable to the strongest known effects as a function of environment \citep[e.g.][]{ramakrishnan_cosmic_2019}. At given peak mass, the bottom quintile in \rl is uncorrelated with the large-scale density field for $M_{peak}\leq\ttt{13.5}\si{\Msolh}$.


Our future work with the \methabb will focus on the three applications already outlined in this paper, the bias of the peaks, their internal density structure, and the structure and geometry of the cosmic web, as represented by the percolating isodensity surface. Regarding assembly bias as a function of \rl and the density structure of peaks, it will be interesting to see how these imprint themselves on the properties of the galaxies residing in the peaks, and whether this can provide further understanding of the connection between the properties of galaxies and their larger scale clustering.

 



\bibliography{phd} 
\bibliographystyle{mnras}

\appendix

\section{Connection to the NFW-Concentration}\label{sec:nfw_conc_def}

  \begin{figure*}
    \centering
    \includegraphics[scale=1.5]{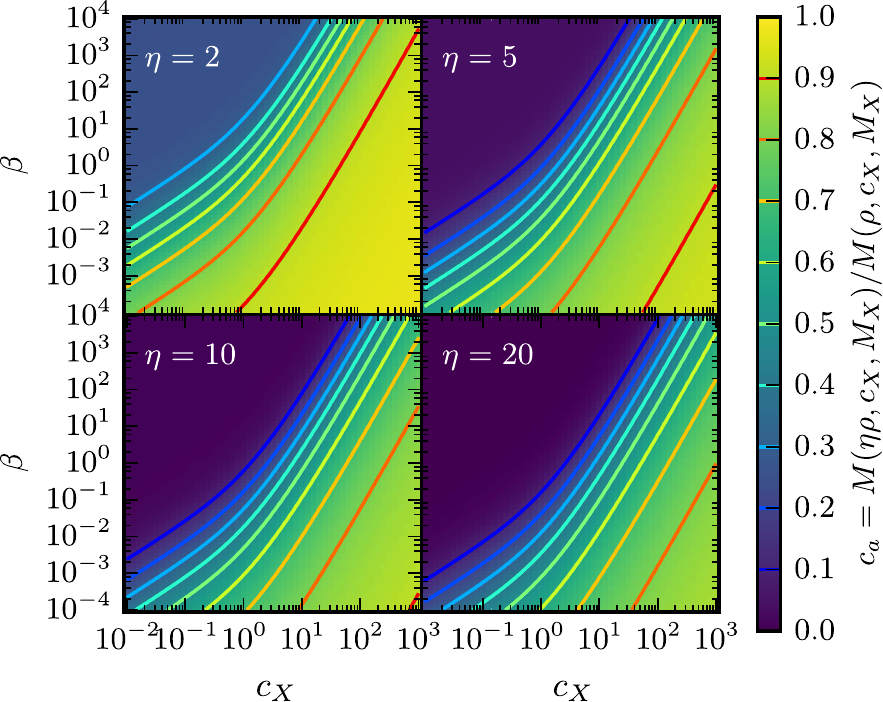}
    \caption[Concentration-Mass Ratio Relation]{The relation between the NFW-concentration $c_X$ and the mass ratio $c_a$ of the contents of two isodensity surfaces with a density ratio $\eta$, where the lower of the isodensity surfaces lies at $\beta$ times the NFW reference overdensity $\delta_X$.}\label{fig:conc_rat}
  \end{figure*}
    
  Another measure of concentration is the well known concentration parameter of the NFW-profile. The bijective nature of the profile function allows us to invert it and obtain the function $r(\rho,c_X,\delta_X)$ with a characteristic overdensity $\delta_X$, corresponding concentration parameter $c_X$ and evaluated at a density $\rho$. The system of reference is set by a choice of $\delta_X$, frequent choices are $\delta_{200c}$ and $\delta_{200m}$, 200 times the critical and mean density, respectively. 
  
  For the following we are interested in a formulation of this radius which does not depend on the choice of $\delta_X$ or the physical scale of the halo. Due to the nature of the profile we only need the ratio of the density $\rho$ and the reference overdensity
  \begin{equation}
  \beta = \frac{\rho}{\delta_X}
  \end{equation}
  which gives us
  \begin{equation}
  \begin{split}
  r\left(\beta, c_X, R_X\right) & = \frac{R_X}{3 c_X}\left(2^{\frac{1}{3}} t(\beta)+2^{-\frac{1}{3}} t(\beta)^{-1}-2\right) \\
  & =\frac{R_X}{c_X}f(\beta),
  \end{split}
  \end{equation}
  with
  \begin{equation}
  t(\beta) = \sqrt[3]{\frac{\beta }{2 \beta+3 \left(\sqrt{12 \beta +81}+9\right)}}.
  \end{equation}

  Using this radius expression we can find the mass inside a bounding density $\rho_b=\beta\delta_X$:
  \begin{equation}
  \begin{split}
    M(\rho_b,c_X,M_X) &= 4\pi \int\limits_0^{r(\beta,c_X,R_X)} \rho_{NFW}(r) r^2 \mathrm{d}r \\
    &= \frac{4\pi}{3}R_X^3\delta_X \frac{1+c_X}{\left(1+c_X\right)\ln\left(1+c_X\right)-c_X}\\ 
    &\,\,\,\,\left[\ln\left(\frac{R_S+R_Xc_X^{-1}f(\beta)}{R_S}\right)-\frac{R_S}{R_S+R_Xc_X^{-1}f(\beta)}\right]\\
    &= M_X \frac{1+c_X}{\left(1+c_X\right)\ln\left(1+c_X\right)-c_X}\\
    &\,\,\,\,\left[\ln\left(1+f(\beta)\right)-\frac{1}{1+f(\beta)}\right]\label{eqn:nfw_mvc}\\
  &= M_X g(\beta,c_X),
  \end{split}
  \end{equation}
  where we used the relation $c_X=R_S/R_X$ for the NFW concentration parameter and 
  \begin{equation}
  M_X = \frac{4\pi}{3}R_X^3\delta_X
  \end{equation}
  for the characteristic mass $M_X$.

  To be able to translate the values found by the mass ratios to concentrations in the NFW framework only need to take the ratio of the masses for two bounding densities with a given $\eta$:
  \begin{equation}
  c_a = \frac{M(\eta\rho,c_X,M_X)}{M(\rho,c_X,M_X)} = \frac{g(\eta\beta\delta_X,c_X)}{g(\beta\delta_X,c_X)} \label{eqn:conc_rat}
  \end{equation}

  Using the relation \eqref{eqn:conc_rat} we can now connect a given mass ratio $c_a$ at a bounding density ratio $\eta$ as discussed in \ref{sec:mass_conc_def} to a NFW-concentration. This is shown in \autoref{fig:conc_rat} for different values of $\eta$. We find that different choices of $\eta$ give different locations of the region of steepest ascent for $c_a$. Depending on the objects of interest certain $\eta$ values should give more precise connections. In the case of unthresholded haloes we want to compare their concentrations defined in relation to a common overdensity, despite them existing over very different mass ranges.

  By fitting such an NFW-profile in density/mass-space, we can find an equivalent concentration value for each object at a given density.
    
\label{lastpage}
\end{document}